\documentclass[12pt,thmsa]{article}
\usepackage{amssymb}
\usepackage{graphicx}

\makeatletter

\topmargin\z@

\def\section{\@startsection{section}{1}{\z@}{-\bigskipamount}{\bigskipamount}
{\bf}}

\newfont{\tts}{cmr5 scaled 1000}

\DeclareSymbolFont{AMSa}{U}{msa}{m}{n}
\DeclareSymbolFont{AMSb}{U}{msb}{m}{n}
\DeclareSymbolFontAlphabet{\mathbb}{AMSb}
\DeclareMathSymbol\square           {\mathord}{AMSa}{"03}

\begin{document}

\title{Discrete phase space - III: The Divergence-free S-matrix elements}
\author{{\small A. Das} \\
%EndAName
\emph{\small Department of Mathematics} \\
\emph{\small Simon Fraser University, Burnaby, B.C. V5A1S6, Canada}}
\date{{\small .}}
\maketitle

In the arena of the discrete phase space and continuous time, the theory of
S-marix is formulated. In the special case of Quantum-Electrodynamics (QED),
the Feynman rules are precisely developed. These rules in the four-momentum
turn out to be identical to the usual QED, \emph{except for the vertex
function}. The new vertex function is given by an infinite series which can
only be treated in an asymptotic approximation at the present time.
Preliminary approximations prove that the second order self-energies of a
fermion and a photon in the discrete model have convergent improper
integrals. In the final section, a sharper asymptotic analysis is employed. 
\emph{It is proved that in case the number of external photon or fermion
lines is at least one, then the S-matrix elements converge in all orders.
Moreover, there are no infra-red divergences in this formulation.}

\section{Introduction}

\qquad It is well known that the Quantum-Electrodynamic (QED) has been
verified to an extraordinary degree of precision by many experiments.
However, it can never explain in its renormalized version the theoretical
values of charge, mass etc. Moreover, it cannot produce in the classical
limit the exact, non-perturbative solution of the coupled Maxwell-Dirac
equations $^{1}$. Therefore, it is worthwhile to investigate the S-matrix,
QED etc. in the discrete phase space and continuous time which involve \emph{%
non-singular Green's functions}$^{2}$, in order to obtain convergent results.

\qquad In the section-II, we compare and contrast the continuous and
discrete methods involving perturbation series and Green's functions for
extremely simple mathematical problems. We conclude that in \emph{linear
problems, both methods incur similar divergence difficulties}. However, in
non-linear problems, the continuous methods encounter divergences whereas
the \emph{discrete methods avoid divergences} in the lower order terms.
Furthermore, the classical potential energy for two particles, as well as
the Green's functions are non-singular in the discrete model. Discussions of
this section offer suggestive insights into the complicated problems of QED
in subsequent sections.

\qquad We assume that the free electromagnetic fields and Dirac fields obey
the \emph{difference-differential} equations of the paper-II. In the
section-III, the interaction picture is introduced and the \emph{relativistic%
} S-matrix is derived in the setting of the discrete phase space and
continuous time.

\qquad In the next section, we introduce the trilinear interaction energy
density of QED in the discrete model. We carefully derive Feynman rules in
the four-momentum space following the difference-differential equations. 
\emph{The four-momentum variables do not have cut off limits in spite the
presence of a characteristic length in the theory}. The only difference with
the usual continuous model is the appearance of the vertex function $\delta
^{\#}$ instead of the usual delta functions $\delta $. The distribution
function $\delta ^{\#}$ is both the joy and the pain of this research
project. It is a joy since $\delta ^{\#}$ is different from $\delta $ (see
the Appendix) and there is a slim chance of obtaining the divergence-free
S-matrix. On the other hand, it is a pain because of the exceedingly
difficult mathematics involved. We have obtained a simple \emph{asymptotic
approximation} $d^{\#}$ for $\delta ^{\#}$ and apply it to the computations
of the next section.

\qquad In the section-V, we apply the Feynman rules for the discrete
phase-space to evaluate the second order fermionic as well as the photon
self energy. \emph{In both cases, the improper integrals converge}. Since
the second order terms are the most dominant terms in the corresponding
S-matrix series, these convergences are welcome news. In section-VI, the
asymptotic analysis of the distribution function $\delta ^{\#}$ and the
infinite series representing an S-matrix element is sharpened considerably.
We investigated an S-matrix element corresponding to the physical process
involving $E_{B}$ external boson lines and $E_{F}$ external fermion lines.
We prove that in the framework of the discrete phase space, \emph{the
S-matrix elements converge in all orders provided} $E_{B}+(3/2)E_{F}>1$.

In the present formalism, the basic physical frame-work remains unchanged.
Only the representations of the quantum mechanics and the quantum field
theory are altered. Moreover, there are \emph{no infra-red divergences.}

\section{Comparison of continuous and discrete methods in simple problems}

\qquad Let us consider a continuous wave function $\hbox{exp}(ikx)$ in a one
dimensional space. It behaves as the following: 
\begin{eqnarray}
&&\lim_{x\rightarrow \infty }\,\hbox{exp}(ikx)\hbox{ is not defined;} 
\nonumber \\
&&\lim_{k\rightarrow \infty }\,\hbox{exp}(ikx)\hbox{ is not defined;} 
\nonumber \\
&&\lim_{L\rightarrow \infty }\,\int_{-L}^{L}|\hbox{exp}(ikx)|^{2}dk\,%
\rightarrow \infty .  \label{eq:one}
\end{eqnarray}

\qquad Now consider a momentum wave function in the discrete model (see
equation II-A.I.6): 
\begin{eqnarray}
\xi _{n}(k) &:&=\frac{(i)^{n}\hbox{exp}(-k^{2}/2)\,H_{n}(k)}{\pi
^{1/4}2^{n/2}\sqrt{n!}},  \nonumber  \label{eq:twotwo} \\
n &\in &\{0,1,2,3,...\}=:\mathbf{N}.
\end{eqnarray}
Here, $H_{n}(k)$ is a Hermite polynomial. In contrast to (\ref{eq:twotwo}),
the wave function $\xi _{n}(k)$ satisfy: 
\begin{eqnarray}
&&\lim_{n\rightarrow \infty }\xi _{n}(k)=\lim_{k\rightarrow \infty }\xi
_{n}(k)=0,  \nonumber \\
&&\lim_{L\rightarrow \infty }\int_{-L}^{L}\left| \xi _{n}(k)\right| ^{2}dk=1.
\end{eqnarray}

\qquad Now let us try to solve an extremely simple, first order, linear,
non-homogeneous ordinary differential equation 
\begin{equation}
\psi ^{\prime }(x)=\phi (x),\qquad x\in \mathbf{R},  \label{eq:twofour}
\end{equation}
by the method of Green's function. Here, $\phi (x)$ is a prescribed
differentiable complex-valued function over \textbf{R}.

\qquad The Green's function for (\ref{eq:twofour}) is given by 
\begin{equation}
G(x-\hat{x}):=(2 \pi i)^{-1} \left(\hbox{C.P.V.}\right)\int_{-\infty}^{%
\infty}k^{-1} \hbox{exp}\left[i k(x-\hat{x})\right]dk  \nonumber
\end{equation}

\begin{eqnarray}
= \frac{1}{2} \frac{(x-\hat{x})}{|(x-\hat{x})|} &\hbox{for}\,\, x\neq\hat{x},
\nonumber \\
0 &\hbox{for}\,\, x=\hat{x};  \nonumber
\end{eqnarray}
\begin{eqnarray}
&\frac{\partial}{\partial x}G(x-\hat{x}) = \delta(x-\hat{x})  \nonumber \\
&\lim_{\hat{x}\rightarrow x} \left|\frac{\partial}{\partial x}G(x-\hat{x})
\right| \rightarrow \infty.  \label{eq:twofive}
\end{eqnarray}

\qquad The general solution of (\ref{eq:twofour}), with help of (\ref
{eq:twofive}), is furnished by

\begin{eqnarray}
\psi (x) &=&\alpha +\left( \hbox{C.P.V.}\right) \int_{-\infty }^{\infty }{%
G(x-\hat{x})\phi (\hat{x})\,d\hat{x}},  \nonumber \\
&=&\alpha +\frac{1}{2}\lim_{L\rightarrow \infty }\left[ \int_{-L}^{x}{\phi (%
\hat{x})\,d}\widehat{{x}}-\int_{x}^{L}{\phi (\hat{x})\,d\hat{x}}\right] ,
\label{eq:twosix}
\end{eqnarray}
where $\alpha $ is an arbitrary complex constant of integration.

\qquad Now we shall solve the corresponding complex, linear, nonhomogeneous 
\emph{difference} equation 
\begin{eqnarray}
\Delta ^{\#}\,\,\psi (n):= &&\frac{1}{\sqrt{2}}\left[ \sqrt{n+1}\,\psi (n+1)-%
\sqrt{n}\,\psi (n-1)\right] =\phi (n),  \nonumber \\
n &\in &\mathbf{N}.  \label{eq:twoseven}
\end{eqnarray}
The Green's function for this problem is furnished by 
\begin{eqnarray}
&&G(n,\hat{n}):=(i)^{-1}\left( \hbox{C.P.V.}\right) \int_{-\infty }^{\infty }%
{(k)^{-1}\xi _{n}(k)\overline{\xi _{\hat{n}}(k)}\,dk},  \nonumber \\
&&\Delta ^{\#}\,\,G(n,\hat{n})=\delta _{n\hat{n}},  \nonumber \\
&&\Delta ^{\#}\,\,G(n,\hat{n})_{|\hat{n}=n}\equiv 1.
\end{eqnarray}
The general solution of (\ref{eq:twoseven}) is given by 
\begin{equation}
\psi (n)=\alpha \xi _{n}(0)-i\sum_{\hat{n}=0}^{\infty }\{\phi (\hat{n}%
)\left[ \left( \hbox{C.P.V.}\right) \int_{-\infty }^{\infty }k^{-1}\xi
_{n}(k)\overline{\xi _{\hat{n}}(k)}dk\right] \},  \label{eq:twonine}
\end{equation}
where $\alpha $ is an arbitrary constant.

\qquad Let us apply the solutions (\ref{eq:twosix}) as well as (\ref
{eq:twonine}) to the momentum eigenfunction problem in quantum mechanics,
namely 
\begin{equation}
P\mathbf{\psi }=e\mathbf{\psi }.  \label{eq:twoonezero}
\end{equation}
Here, we assume that the parameter $e\neq 0$ is sufficiently small. In the
continuous (Schroedinger) representation of quantum mechanics (\ref
{eq:twoonezero}) yields the differential equation 
\begin{equation}
\psi ^{\prime }(x)=ie\psi (x),\,\,\,x\in R,\,\,\,\psi (x)\neq 0.
\label{eq:twooneone}
\end{equation}
The general solution of the above equation is obviously given by 
\begin{equation}
\psi (x)=\alpha \,\hbox{exp}(iex)=\alpha \left[ 1+iex+\frac{(iex)^{2}}{2}%
\right] +O(e^{3}).  \label{eq:twoonetwo}
\end{equation}
Here, $\alpha $ is an arbitrary, non-zero, complex constant.

\qquad Now we try to solve (\ref{eq:twooneone}) by the perturbative
expansion and the method of Green's function in (\ref{eq:twosix}).
Substituting an expansion 
\begin{equation}
\psi (x)=\sum_{j=0}^{\infty }\left( ie\right) ^{j}\psi _{j}(x)
\label{eq:twoonethree}
\end{equation}
into (\ref{eq:twooneone}), we derive an infinite string of differential
equations: 
\begin{equation}
\psi _{0}^{\prime }(x)=0,\psi _{1}^{\prime }(x)=\psi _{0}(x),\psi
_{2}^{\prime }(x)=\psi _{1}(x),..,\psi _{j}^{\prime }(x)=\psi _{j-1}(x),...\
.  \label{eq:twoonefour}
\end{equation}
Solving the first equation we obtain 
\begin{equation}
\psi _{0}(x)=\alpha ,  \label{eq:twoonefive}
\end{equation}
where $\alpha \neq 0$ is otherwise arbitrary. In solutions of other
equations, we shall ignore arbitrary constants. Solutions of the next two
equations in (\ref{eq:twoonefour}) by the method of (\ref{eq:twosix}) are
listed below. 
\begin{eqnarray}
\psi _{1}(x) &=&\left( \frac{\alpha }{2}\right) \left[ \lim_{L\rightarrow
\infty }\left( \int_{-L}^{x}d\hat{x}-\int_{x}^{L}d\hat{x}\right) \right] 
\nonumber \\
&=&\alpha x,  \nonumber \\
\psi _{2}(x) &=&\left( \frac{\alpha }{2}\right) \left[ \lim_{L\rightarrow
\infty }\left( \int_{-L}^{x}{\hat{x}\,d\hat{x}}-\int_{x}^{L}{\hat{x}\,d\hat{x%
}}\right) \right]  \nonumber \\
&=&\left( \frac{\alpha }{2}\right) x^{2}-\left( \frac{\alpha }{2}\right)
\left[ \lim_{L\rightarrow \infty }(L^{2})\right] .  \label{eq:twoonesix}
\end{eqnarray}
In the second order solution $\psi _{2}(x)$ we encounter a \emph{divergent
term!} Ignoring this term (``renormalizing''), by (\ref{eq:twoonefive}) and (%
\ref{eq:twoonesix}) we can recover the first three terms of the expansion in
the RHS of (\ref{eq:twoonetwo}).

\qquad Now, we shall try to solve the \emph{same} problem (\ref
{eq:twoonezero}) with the discrete representation of quantum mechanics. The
corresponding difference equation and the exact general solution are
provided by 
\begin{eqnarray}
\Delta ^{\#}\psi (n) &=&ie\psi (n),  \nonumber \\
\psi (n)=\alpha \xi _{n}(e) &=&\alpha \left[ \xi _{n}(0)+e\xi _{n}^{\prime
}(0)+\frac{e^{2}}{2}\xi _{n}^{\prime \prime }(0)\right] +O(e^{3}),  \nonumber
\\
\alpha &\neq &0.  \label{eq:twooneseven}
\end{eqnarray}
Using a perturbative expansion 
\begin{equation}
\psi (n)=\sum_{j=0}^{\infty }\left( ie\right) ^{j}\psi _{j}(n),
\label{eq:twooneeight}
\end{equation}
the difference equation in (\ref{eq:twooneseven}) yields the following
infinite string of difference equations: 
\begin{equation}
\Delta ^{\#}\psi _{0}(n)=0,\Delta ^{\#}\psi _{1}(n)=\psi _{0}(n),..,\Delta
^{\#}\psi _{j}(n)=\psi _{j-1}(n),...  \label{eq:twoonenine}
\end{equation}
The first of the equations (\ref{eq:twoonenine}) is solved by (see
Appendix-I of paper-II) 
\begin{eqnarray}
\psi _{0}(n) &=&\alpha \xi _{n}(0),  \nonumber \\
\psi _{0}(2n) &=&\alpha \xi _{2n}(0),  \nonumber \\
\psi _{0}(2n+1) &=&\alpha \xi _{2n+1}(0)\equiv 0.  \label{eq:twotwozero}
\end{eqnarray}
The second and third equations of (\ref{eq:twoonenine}) can be solved using (%
\ref{eq:twonine}) and (\ref{eq:twotwozero}). Ignoring arbitrary constants,
these solutions are: 
\begin{eqnarray}
\psi _{1}(n) &=&-i\alpha \left[ \left( \hbox{C.P.V.}\right) \int_{-\infty
}^{\infty }{k^{-1}\xi _{n}(k)\delta (k)\,dk}\right] ,  \nonumber \\
\psi _{1}(2n) &\equiv &0,  \nonumber \\
\psi _{1}(2n+1) &=&-i\alpha \{\lim_{k\rightarrow \infty }k^{-1}\left[ \xi
_{2n+1}(k)-\xi _{2n+1}(0)\right] \}  \nonumber \\
&=&-i\alpha \xi _{2n+1}^{\prime }(0),  \nonumber \\
\psi _{2}(2n) &=&-\alpha \{\left( \hbox{C.P.V}\right) \int_{-\infty
}^{\infty }{k^{-1}\xi _{2n}(k)\left[ \sum_{\hat{n}=0}^{\infty }\xi _{2\hat{n}%
+1}^{\prime }(0)\overline{\xi _{2\hat{n}+1}(k)}\right] dk}\}  \nonumber \\
&=&-\alpha \xi _{2n}^{\prime \prime }(0)-\lim_{k\rightarrow 0}\left[
k^{-2}\xi _{2n}(k)\right] ,  \nonumber \\
\psi _{2}(2n+1) &\equiv &0.  \label{eq:twotwoone}
\end{eqnarray}
(In deriving equations (\ref{eq:twotwoone}), we have used 
\begin{equation}
\sum_{-\hat{n}=0}^{\infty }\xi _{2\hat{n}+1}^{\prime }(0)\overline{\xi _{2%
\hat{n}+1}(k)}=-\delta ^{\prime }(k),\int_{-\infty }^{\infty }f(k)\delta
^{\prime }(k)\,dk=-f^{\prime }(k)  \nonumber
\end{equation}
etc.) The RHS of the equation for $\psi _{2}(2n)$ has a \emph{divergent
constant}. Ignoring it and using equations (\ref{eq:twooneeight}), (\ref
{eq:twotwozero}) and (\ref{eq:twotwoone}) we can recover the three terms of
the RHS of the series in equation (\ref{eq:twooneseven}). So, we discover
that the solution of the \emph{linear} quantum mechanical problem in (\ref
{eq:twoonezero}) with perturbative expansion and Green's functions produces 
\emph{exactly similar divergence difficulties in the continuous or the
discrete representation}.

\qquad Now let us investigate a simple non-linear toy model by the two
methods. Consider a first order non-linear ordinary differential equation: 
\begin{equation}
\psi ^{\prime }(x)=e\left[ \psi (x)\right] ^{2}.
\end{equation}
The exact general solution is given by 
\begin{equation}
\psi (x)=\alpha \left[ 1-e\alpha x\right] ^{-1},  \label{eq:nonsol}
\end{equation}
where $\alpha $ is an arbitrary complex constant. In the case where $\alpha $
is a non-zero real constant, the solution in (\ref{eq:nonsol}) has a \emph{%
singularity} at $x=(e\alpha )^{-1}$. In case $|\alpha x|<1$, we can elicit
from (\ref{eq:nonsol}) a series expansion 
\begin{equation}
\psi (x)=\alpha \sum_{j=0}^{\infty }(e\alpha x)^{j}.  \label{eq:twotwofour}
\end{equation}
A perturbative expansion 
\begin{equation}
\psi (x)=\sum_{j=0}^{\infty }(e)^{j}\psi _{j}(x)  \label{eq:twotwofive}
\end{equation}
leads to the following string of differential equations: 
\begin{equation}
\psi _{0}^{\prime }(x)=0,\psi _{1}^{\prime }(x)=[\psi _{0}(x)]^{2},\psi
_{2}^{\prime }(x)=2\psi _{0}(x)\psi _{1}(x),...\,\,\,\,\,.
\label{eq:twotwosix}
\end{equation}
Using the solution (\ref{eq:twosix}) involving the Green's function and
ignoring arbitrary constants after the first solution, we obtain 
\begin{eqnarray}
\psi _{0}(x) &=&\alpha ,\psi _{1}(x)=\alpha ^{2}x,  \nonumber \\
\psi _{2}(x) &=&\alpha ^{3}\lim_{L\rightarrow \infty }\left[ \int_{-L}^{x}{%
\hat{x}\,d\hat{x}}-\int_{x}^{L}{\hat{x}\,d\hat{x}}\right]   \nonumber \\
&=&\alpha ^{3}x^{2}-\alpha ^{3}\lim_{L\rightarrow \infty }L^{2}.
\label{eq:twotwoseven}
\end{eqnarray}
Ignoring the \emph{divergent term} in the last equation, using equations (%
\ref{eq:twotwofive}) and (\ref{eq:twotwoseven}), we recover the first three
terms in the expansion of (\ref{eq:twotwofour}).

\qquad The corresponding non-linear \emph{difference} equation is furnished
by 
\begin{equation}
\Delta ^{\#}\psi (n)=e[\psi (n)]^{2}.  \label{eq:twotwoeight}
\end{equation}
With a perturbative expansion 
\begin{equation}
\psi (n)=\sum_{j=0}^{\infty }e^{j}\psi _{j}(n),
\end{equation}
the equation (\ref{eq:twotwoeight}) implies that 
\begin{equation}
\Delta ^{\#}\psi _{0}(n)=0,\Delta ^{\#}\psi _{1}(n)=[\psi
_{0}(n)]^{2},\Delta ^{\#}\psi _{0}(n)=2\psi _{0}(n)\psi _{1}(n),...::.
\label{eq:twothreeoh}
\end{equation}
The solutions of these equations by the method of Green's function in (\ref
{eq:twonine}) lead to the following expressions: 
\begin{eqnarray}
\psi _{0}(n) &=&\alpha \xi _{n}(0),  \nonumber \\
\psi _{1}(2n) &=&-i\alpha ^{2}\sum_{\hat{n}=0}^{\infty }\{[\xi _{2\hat{n}%
}(0)]^{2}\left[ \left( \hbox{C.P.V.}\right) \int_{-\infty }^{\infty }{%
k^{-1}\xi _{2n}(k)\overline{\xi _{2\hat{n}}(k)}\,dk}\right] \}  \nonumber \\
&\equiv &0,
\end{eqnarray}
\begin{eqnarray}
\psi _{1}(2n+1) &=&-i\alpha ^{2}\left( \hbox{C.P.V.}\right) \int_{-\infty
}^{\infty }{\{k^{-1}\xi _{2n+1}(k)\left[ \sum_{\hat{n}=0}^{\infty }[\xi _{2%
\hat{n}}(0)]^{2}\overline{\xi _{2\hat{n}}(k)}\right] dk}  \nonumber \\
&\simeq &-i\alpha ^{2}\left( \hbox{C.P.V.}\right) \int_{-\infty }^{\infty }{%
(2\pi k)^{-1}\xi _{2\hat{n}}(k)[4\sqrt{|k|}\delta (k)+}  \nonumber \\
&&{|k]|^{-1/2}dk}.
\end{eqnarray}
(Here we have used A.18.) In the last integral \emph{there are neither
``ultraviolet'' nor ``infrared'' divergences}.

\qquad The third equation in (\ref{eq:twothreeoh}) yields the solution 
\begin{eqnarray}
\psi _{2}(n) &=&-2\alpha ^{3}\sum_{\hat{n}=0}^{\infty }\sum_{n=0}^{\infty
}\{[\xi _{2\hat{n}}(0)][\xi _{n}(0)]\left[ \left( \hbox{C.P.V.}\right)
\int_{-\infty }^{\infty }\int_{-\infty }^{\infty }(\hat{k}k)^{-1}\right. 
\nonumber \\
&&\left. \xi _{2\hat{n}}(\hat{k})\overline{\xi _{2n}(\hat{k})}\xi _{n}(k)%
\overline{\xi _{2\hat{n}}(\hat{k})}\,d\hat{k}\,dk\right] \}  \nonumber \\
&\equiv &0.  \label{eq:twothreetwo}
\end{eqnarray}
In the RHS of equation (\ref{eq:twothreetwo}), the (C.P.V.) integral \emph{%
converges} to zero since the integrand is \emph{odd} with respect to the
variable $\hat{k}$. Therefore we conclude that for a \emph{non-linear
problem, convergence is more likely in the discrete method} compared to the
continuous case.

\qquad No we shall discuss the \emph{vertex function} in the one dimensional
momentum space in the case of a \emph{trilinear} interaction term. In the
usual continuous case, it is given by 
\begin{equation}
(2\pi )^{-1}\left( \hbox{C.P.V.}\right) \int_{-\infty }^{\infty }\hbox{exp}%
\left[ i\left( p-q-k\right) x\right] dx=\delta (p-q-k).
\label{twothreethree}
\end{equation}
The above equation indicates a \emph{sharp} conservation of momenta or wave
numbers for three particles. In case functions are defined in the half line $%
x>0$, the corresponding vertex function is furnished by$^{3}$%
\begin{equation}
(2\pi )^{-1}\int_{0}^{\infty }\hbox{exp}\left[ i\left( p-q-k\right) x\right]
dx=\frac{1}{2}\{\delta (p-q-k)+\frac{i}{2}\frac{sgn(p-q-k)}{|p-k-k|}\}.
\label{twothreefour}
\end{equation}
In the RHS, there is one term indicating the \emph{sharp} conservation and
another is representing one \emph{soft} conservation. In the field theory in
an one dimensional lattice space, the corresponding vertex function is$^{4}$%
\begin{equation}
(2\pi )^{-1}\sum_{n=-\infty }^{\infty }\hbox{exp}[i(p-q-k)n]=\sum_{j=-\infty
}^{\infty }\delta (p-q-k+2\pi j).  \label{twothreefive}
\end{equation}
The above indicates a denumerably infinite sharp conservation of wave
numbers. In a single discrete phase plane, the corresponding vertex function
is specified by (see equations A.1): 
\[
\delta ^{\#}(p,-q,-k):=\sum_{n=0}^{\infty }\xi _{n}(p)\overline{\xi _{n}(q)}%
\,\overline{\xi _{n}(k)}\simeq d^{\#}(p,-q,-k).
\]
In the special case of $p=0$, it is explicitly furnished by (see A.18): 
\begin{eqnarray}
d^{\#}(0,-q,-k) &=&\frac{1}{\pi }\{\sqrt{|q+k|}\delta (q+k)+\sqrt{|q-k|}%
\delta (q-k)  \nonumber \\
&+&\frac{1}{4}\left[ |q+k|^{-1/2}+|q-k|^{-1/2}\right] \}.
\label{eq:twothreeseven}
\end{eqnarray}
The RHS of the above equation demonstrates \emph{two sharp} as well as \emph{%
two soft} conservations. Moreover, \emph{spontaneous reflections} of momenta
(on lattice points) are allowed. (These reflections are analogous to Bragg
reflections on the lattice planes of a crystal.)

\qquad Now we investigate the \emph{potential equation} in our discrete
model. It is given by: 
\begin{equation}
\delta ^{ab}\Delta _{a}^{\#}\Delta _{b}^{\#}V(\mathbf{n})=0,\ (\mathbf{n}%
):=(n^{1},n^{2},n^{3})\in \mathbf{N}^{3}.  \label{eq:twothreeeight}
\end{equation}
The corresponding Green's function (which is the potential energy between
two unit charges) satisfy: 
\begin{eqnarray}
G\left( \mathbf{n},\mathbf{\hat{n}}\right) &=&\int_{R^{3}}{(\mathbf{k}\cdot 
\mathbf{k})}^{-1}{\left[ \prod_{a=1}^{3}\xi _{n^{a}}(k_{a})\overline{\xi _{%
\hat{n}^{a}}(k_{a})}\right] }d^{3}\mathbf{k},  \nonumber \\
\lim_{\mathbf{\hat{n}}\rightarrow \mathbf{n}}\left| G(\mathbf{n},\mathbf{%
\hat{n}})\right| &<&\infty .  \label{eq:twothreenine}
\end{eqnarray}
The convergences of the integrals above are due to the facts that $%
H_{n^{\alpha }}(k_{\alpha })$ is a polynomial and there is a decaying weight
factor $\hbox{exp}[-(\mathbf{k}\cdot \mathbf{k})]$ in products of $\xi
_{n^{a}}(k_{a})$.

\qquad Now we consider the Green's functions (equation (II-A.II.1.B)) for
the difference-differential Klein-Gordon equation. The coincidence limits
are provided by: 
\begin{eqnarray}
\Delta_{(a)}(\mathbf{n}, t; \mathbf{n}, t; \mu)&=&
(2\pi)^{-1}\pi^{-3/2}\int_{R^3}\lbrace\hbox{exp}[-k_{1}^{2}-k_{2}^{2}
-k_{3}^{2}] \left[\prod_{j=1}^{3} \frac{\left[H_{n^j}(k_{j})\right]^2}{%
2^{n^j}(n^j)!}\right]  \nonumber \\
&\times& \left[\int_{C_{(a)}} \left(\eta^{\alpha\beta}k_{\alpha}k_{\beta} +
\mu^2 \right)^{-1}dk^4\right]\rbrace d^3\mathbf{k}.  \label{eq:twofourzero}
\end{eqnarray}
Because of the weight factor $\hbox{exp}[-k_{1}^{2}-k_{2}^{2}-k_{3}^{2}]$,
the integral on the RHS of (\ref{eq:twofourzero}) \emph{converges}. All of
the Green's functions for free fields in our discrete model are \emph{%
non-singular}.

\qquad The discussions of this present section will provide valuable
insights into the complicated topics of the following sections.

\section{The interaction picture and the S-matrix}

\qquad We shall follow the same notations as in the previous papers I and
II. The equations of the relativistic quantum fields are expressed \emph{%
exclusively} by the partial \emph{difference-differential} equations in this
paper. We shall now derive the S-matrix for interacting fields in the
sequel. In the interaction picture, the time-evolution of a Hilbert space
(state) vector $\left| \Psi _{I}(t)\rangle \right. $ (representing a many
particle system) is governed by the differential equation$^{5}$ 
\begin{equation}
i\partial _{t}\left| \Psi _{I}(t)\rangle \right. =H_{I}(t)\left| \Psi
_{I}(t)\rangle \right. .  \label{eq:threeone}
\end{equation}
Here, $H_{I}(t)$ represents the hermitian operator corresponding to the 
\emph{total interaction energy} at the instant $t$. We can express $H_{I}(t)$
in terms of the \emph{interaction energy density operator} $\hbox{$\cal{H}$}%
_{I}(\mathbf{n},t)$ by the triple sum: 
%%\begin{mathletters}
\begin{eqnarray}
H_{I}(t):= &&{\sum_{\mathbf{n}=\mathbf{0}}^{\infty }}^{(3)}\hbox{$\cal{H}$}%
_{I}(\mathbf{n},t),  \label{eq:threetwo} \\
\hbox{$\cal{H}$}_{I}(t) &=&-\hbox{$\cal{L}$}_{I}(\mathbf{n},t).
\label{eq:threetwob}
\end{eqnarray}
Here, $\hbox{$\cal{L}$}_{I}(\mathbf{n},t)$ stands for the interaction
Lagrangian density and it is a \emph{relativistic invariant} operator. (The
equation (\ref{eq:threetwob}) holds for most of the useful cases.) A
necessary micro-causality requirement is 
%%\end{mathletters}
\begin{equation}
\lbrack \hbox{$\cal{H}$}_{I}(\mathbf{n},t),\hbox{$\cal{H}$}_{I}(\mathbf{\hat{%
n}},t)]\equiv 0\,\,\,\,\hbox{ for }\mathbf{n}\neq \mathbf{\hat{n}}.
\label{eq:threethree}
\end{equation}
Usually, the interaction energy density operator $\hbox{$\cal{H}$}_{I}(%
\mathbf{n},t)$ has an overall multiplier which is \emph{small}. It is
customary to solve (\ref{eq:threeone}) by \emph{perturbative} series
involving this small parameter. Moreover, the S-matrix is the operator which
takes a prescribed initial state $|i\rangle $ into another prescribed final
state $|f\rangle $ consistent with the evolution equation (\ref{eq:threeone}%
). The perturbative series expansion for the S-matrix is furnished by$^{6}$: 
\begin{eqnarray}
S &=&I+\sum_{j=1}^{\infty }S_{j}=I+\sum_{j=1}^{\infty }[(-i)^{j}/j!]{\sum_{%
\mathbf{n}_{1}=\mathbf{0}}^{\infty }}^{(3)}\int_{\mathbf{R}}dt_{1}{\sum_{%
\mathbf{n}_{2}=\mathbf{0}}^{\infty }}^{(3)}\int_{\mathbf{R}}dt_{2}... 
\nonumber \\
&&{\sum_{\mathbf{n}_{j}=\mathbf{0}}^{\infty }}^{(3)}\int_{\mathbf{R}%
}dt_{j}\{T[\hbox{$\cal{H}$}_{I}(\mathbf{n}_{1},t_{1})\,\hbox{$\cal{H}$}_{I}(%
\mathbf{n}_{2},t_{2})...\hbox{$\cal{H}$}_{j}(\mathbf{n}_{j},t_{j})]\}.
\label{eq:threefour}
\end{eqnarray}
Here, $T$ stands for Wick's \emph{time ordering operator}. There is another
operator $N$ called \emph{normal ordering}$^{6}$. It arranges creation
operators to the left of annihilation operators. Furthermore, there is still
another operation called \emph{contraction} between two operators$^{7}$ and
it is defined by: 
\begin{equation}
A(\mathbf{n},t)B(\mathbf{\hat{n}},\hat{t}):=T[A(\mathbf{n},t)B(\mathbf{\hat{n%
}},\hat{t})]-N[A(\mathbf{n},t)B(\mathbf{\hat{n}},\hat{t})].
\label{eq:threefive}
\end{equation}
We can extract from the commutation and anti-commutation relations
(II-3.10B), (II-4.9B), (II-5.11Bi-vii) and the linear relationships
(II-A.II.4B), (II-A.II.8c), the following examples of contractions: 
\begin{eqnarray}
A_{\mu }(\mathbf{n},t)A_{\nu }(\mathbf{\hat{n}},\hat{t}) &=&-i\eta _{\mu \nu
}D_{F}(\mathbf{n},t;\mathbf{\hat{n}},\hat{t})\,I,  \label{eq:threesix} \\
\Psi (\mathbf{n},t)\Psi (\mathbf{\hat{n}},\hat{t}) &=&\tilde{\Psi}(\mathbf{n}%
,t)\tilde{\Psi}(\mathbf{\hat{n}},\hat{t})\equiv 0,  \nonumber \\
\Psi (\mathbf{n},t)\tilde{\Psi}(\mathbf{\hat{n}},\hat{t}) &=&-\tilde{\Psi}(%
\mathbf{\hat{n}},\hat{t})\Psi (\mathbf{n},t)=iS_{F}(\mathbf{n},t;\mathbf{%
\hat{n}},\hat{t};m)\,I,  \nonumber \\
\Psi (\mathbf{n},t)A_{\mu }(\mathbf{\hat{n}},\hat{t}) &=&\tilde{\Psi}(%
\mathbf{n},t)A_{\mu }(\mathbf{\hat{n}},\hat{t})\equiv 0.  \nonumber
\end{eqnarray}
The Green's functions $\Delta _{F}(..)$, $D_{F}(..)$, and $S_{F}(..)$ are
all defined in Appendix-II of paper-II. (We should mention that these
Green's functions are analogues of the causal Green's functions of
Stuckelberg in the +2 signature and differ from the corresponding
Feynman-Dyson propagators by a factor ($-i2$).)

\qquad Now, we shall state the Wick's theorem$^{7}$ on the decomposition of
a chronological product in the arena of the discrete phase space and
continuous time. It can be succinctly stated as: 
\begin{eqnarray}
&&T[A(\mathbf{n}_{1},t_{1})B(\mathbf{n}_{2},t_{2})C(\mathbf{n}%
_{3},t_{3}).....\,\,J(\mathbf{n}_{j},t_{j})]  \nonumber \\
&=&N[A(\mathbf{n}_{1},t_{1})B(\mathbf{n}_{2},t_{2})C(\mathbf{n}%
_{3},t_{3}).....\,\,J(\mathbf{n}_{j},t_{j})]  \nonumber \\
&&+N\{[A(\mathbf{n}_{1},t_{1})B(\mathbf{n}_{2},t_{2})C(\mathbf{n}%
_{3},t_{3}).....\,\,J(\mathbf{n}_{j},t_{j})]+%
\hbox{all other single
contractions}\}  \nonumber \\
&&+....  \nonumber \\
&&+N\{[A(\mathbf{n}_{1},t_{1})B(\mathbf{n}_{2},t_{2})C(\mathbf{n}%
_{3},t_{3}).....\,\,J(\mathbf{n}_{j},t_{j})]+%
\hbox{all other double
contractions}\}  \nonumber \\
&&+....  \label{eq:threeseven} \\
&&+N\{\hbox{terms with maximal number of contractions}\}.  \nonumber
\end{eqnarray}
(This theorem is provable by induction.)

\section{Feynman rules of Q.E.D. in discrete phase-space and continuous time}

\qquad We choose the interaction density as 
\begin{equation}
\hbox{$\cal{H}$}_{I}(\mathbf{n},t):=-ieN[\tilde{\Psi}(\mathbf{n}%
,t)\gamma^{\mu} \Psi(\mathbf{n},t) A_{\mu}(\mathbf{n},t)].
\label{eq:fourone}
\end{equation}
Here, the parameter $|e|=\sqrt{4\pi/137}$ is a small positive number. The
interaction term in (\ref{eq:fourone}) is derived from the principle of 
\emph{minimal electromagnetic interaction}. According to this principle, the
difference and differential operators $\Delta_{a}^{\#}$, $\partial_{t}$ in
the free Lagrangian density (II-5.3) are replaced by $\Delta_{a}^{%
\#}-ieA_{a}(\mathbf{n},t)$ and $\partial_{t}-ieA_{4}(\mathbf{n},t)$
respectively. Moreover, from the discussions in section-IV of paper-I, it is
amply clear that the interaction energy density in (\ref{eq:fourone}) is a 
\emph{relativistic invariant}. Furthermore, the micro-causality condition
(3.3) is satisfied by interaction (\ref{eq:fourone}) due to equations
(II-3.10B), (II-4.9B), and (II-5.11B i-vii).

\qquad The equations (3.4) and (\ref{eq:fourone}) yield for the S-matrix: 
\begin{eqnarray}
S=I+\sum_{j=1}^{\infty}S_{j} = I+\sum_{j=1}^{\infty}[(-e)^{j}/j!] {\sum_{%
\mathbf{n_1}=\mathbf{0}}^{\infty}}^{(3)} \int_{\mathbf{R}}dt_{1} ... {\sum_{%
\mathbf{n_j}=\mathbf{0}}^{\infty}}^{(3)} \int_{\mathbf{R}}dt_{j}  \nonumber
\\
T\lbrace N[\tilde{\Psi}(\mathbf{n}_{1},t_1) \gamma^{\mu_1} \Psi(\mathbf{n}%
_{1},t_1) A_{\mu_1}(\mathbf{n}_{1},t_1)] ...  \nonumber \\
N [\tilde{\Psi}(\mathbf{n}_{j},t_j) \gamma^{\mu_j} \Psi(\mathbf{n}_{j},t_j)
A_{\mu_j}(\mathbf{n}_{j},t_j)]\rbrace .  \label{eq:fourtwo}
\end{eqnarray}
The R.H.S. of (\ref{eq:fourtwo}) is a \emph{relativistic invariant operator}%
. Let us consider the second order term $S_2$ in (\ref{eq:fourtwo}) for the
sake of simplicity. By the Wick's decomposition (\ref{eq:threeseven}) and
equations (\ref{eq:threesix}i-v), $S_2$ can be reduced to the following sum$%
^5$: 
\begin{eqnarray}
S_{2}=(e^{2}/2){\sum_{\mathbf{n_{1}}=\mathbf{0}}^{\infty}}^{(3)} \int_{%
\mathbf{R}} dt_{1} {\sum_{\mathbf{n_{2}}=\mathbf{0}}^{\infty}}^{(3)}\int_{%
\mathbf{R}} dt_{2} [\sum_{A=1}^{8}\Gamma_{(A)}(\mathbf{n_{1}}, t_{1}; 
\mathbf{n_{2}}, t_{2})]  \label{eq:fourthree} \\
=:\sum_{A=1}^{8}S_{2(A)} .  \nonumber
\end{eqnarray}
Two of the operators $\Gamma_{(A)}(..)$ are explicitly furnished below as: 
\begin{eqnarray}
\Gamma_{(5)}(\mathbf{n_1}, t_{1}; \mathbf{n_2}, t_{2}) &=& -N[\gamma^{\mu}
\Psi(\mathbf{n_1}, t_{1}) S_{F}(\mathbf{n_2}, t_{2};\mathbf{n_1}, t_{1}; m)
\eta_{\mu\nu} D_{F}(\mathbf{n_2}, t_{2};\mathbf{n_1}, t_{1}) \tilde{\Psi}(%
\mathbf{n_2}, t_{2})\gamma^{\nu}],  \nonumber \\
\Gamma_{(6)}(\mathbf{n_1}, t_{1}; \mathbf{n_2}, t_{2}) &=& \Gamma_{(5)}(%
\mathbf{n_2}, t_{2};\mathbf{n_1}, t_{1}).  \label{eq:fourfour}
\end{eqnarray}
The second order terms $S_{2(5)}=S_{2(6)}$ contribute towards the
self-energy of an electron.

\qquad Let us now work out the matrix entry $\langle f|\Gamma
_{(5)}(..)|i\rangle $ for the initial state of one electron $|i\rangle
:=\alpha _{r}^{\dagger }(\mathbf{p})|\Psi _{0}\rangle $ and the final state
of one electron $|f\rangle :=\alpha _{s}^{\dagger }(\widehat{\mathbf{p}}%
)|\Psi _{0}\rangle $.

\qquad Using equations (II-5.7Bii), (II-5.8), (II-5.10), (II-A.II.5B), and
(II-A.II.6B), the operator $\Gamma _{(5)}(..)$ in (\ref{eq:fourfour}) yields 
\begin{eqnarray}
\langle f|\Gamma _{(5)}(..)|i\rangle &=&m[E(\mathbf{\hat{p}})E(\mathbf{p}%
)]^{-1/2}\tilde{\mathbf{u}}_{s}(\widehat{\mathbf{p}})\gamma ^{\mu
}\{\lim_{\epsilon \rightarrow 0_{+}}\int_{\mathbf{R}^{4}}\int_{\mathbf{R}%
^{4}}d^{4}k\,d^{4}q  \nonumber \\
&&\left[ \prod_{b=1}^{3}\overline{\xi _{n_{2}^{j}}(\hat{p}_{b})}\xi
_{n_{2}^{j}}(k_{b})\xi _{n_{2}^{j}}(q_{b})\right] (2\pi )^{-1}\exp [i(\hat{E}%
+k_{4}+q_{4})t_{2}]\eta _{\mu \nu }  \nonumber \\
&&(k^{\alpha }k_{\alpha }-i\epsilon )^{-1}(i\gamma ^{\rho }q_{\rho
}-mI)(q^{\beta }q_{\beta }+m^{2}-i\epsilon )^{-1}\gamma ^{\nu } \\
&&\left[ \prod_{a=1}^{3}\xi _{n_{1}^{a}}(p_{a})\overline{\xi
_{n_{1}^{a}}(k_{a})}\overline{\xi _{n_{1}^{a}}(q_{a})}\right] (2\pi
)^{-1}\exp [-i(E+k_{4}+q_{4})t_{1}]\}\mathbf{u}_{r}(\mathbf{p}),  \nonumber
\\
d^{4}q &:&=d^{3}\mathbf{q}\,dq^{4}=-d^{3}\mathbf{q}dq_{4}.  \nonumber
\label{eq:fourfive}
\end{eqnarray}

\qquad We shall now define a new distribution function (see the Appendix) by
the triple sum 
\begin{eqnarray}
\delta_{3}^{\#}(\mathbf{p,q,k}):={\sum_{\mathbf{n} = \mathbf{0}}^{\infty}}%
^{(3)} [\prod_{j=1}^{3} \xi_{n^{j}}(p_{j}) \xi_{n^{j}}(q_{j})
\xi_{n^{j}}(k_{j})]  \nonumber \\
= \sum_{n^{1}=0}^{\infty} \sum_{n^{2}=0}^{\infty} \sum_{n^{3}=0}^{\infty}
[\xi_{n^{1}}(p_{1}) \xi_{n^{1}}(q_{1})\xi_{n^{1}}(k_{1})]  \nonumber \\
\left[\xi_{n^{2}}(p_{2}) \xi_{n^{2}}(q_{2})\xi_{n^{2}}(k_{2})\right]
[\xi_{n^{3}}(p_{3}) \xi_{n^{3}}(q_{3})\xi_{n^{3}}(k_{3})].
\label{eq:foureight}
\end{eqnarray}
Therefore, by the equations (\ref{eq:fourthree}), (\ref{eq:fourfive}), and (%
\ref{eq:foureight}), we obtain the second order contributions for a
fermionic self-energy as 
\begin{eqnarray}
\langle f | S_{2} |i\rangle = \langle f | S_{2(5)} + S_{2(6)} | i \rangle =
(2!) \langle f | S_{2(5)} |i\rangle  \nonumber \\
=e^{2} m[E(\hat{\mathbf{p}})E(\mathbf{p})]^{-1/2} \tilde{\mathbf{u}}_{s}(\hat%
\mathbf{p}) \lbrace \lim_{\epsilon\rightarrow 0_{+}} \int_{\mathbf{R^{4}}%
}\int_{\mathbf{R^{4}}} d^{4}q\,d^{4}k \overline{\delta^{\#}_{3}(\hat{\mathbf{%
p}},-\mathbf{q},-\mathbf{k})}  \nonumber \\
\delta (\hat{E}-q^{4}-k^{4})\gamma^{\mu}(q_{\alpha}q^{\alpha} + m^{2}
-i\epsilon)^{-1} (i\gamma^{\rho}q_{\rho}-mI) \eta_{\mu\nu}
(k^{\beta}k_{\beta}-i\epsilon)^{-1} \gamma^{\nu}  \label{eq:fournine} \\
\delta^{\#}_{3}(\mathbf{p},-\mathbf{q},-\mathbf{k})\delta
(q^{4}+k^{4}-E)\rbrace \mathbf{u}_{r}(\mathbf{p}).  \nonumber
\end{eqnarray}

\qquad A graphic way to represent the right hand side of the equation (\ref
{eq:fournine}) is by Fig.1.

\begin{figure}[ht]
\begin{center}
\includegraphics[bb=230 200 372 602, scale=0.4, clip, keepaspectratio=true]{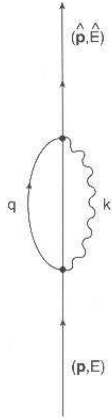}
\caption{{\small The second order fermion self-energy graph in the momentum space}}
\label{fig:one}
\end{center}
\end{figure}

\qquad This expression and other matrix elements give rise to
table -1 of Feynman Rules.

\begin{figure}[ht]
\begin{center}
\includegraphics[bb=46 131 555 664, angle=0.90, scale=0.62, clip, keepaspectratio=true]{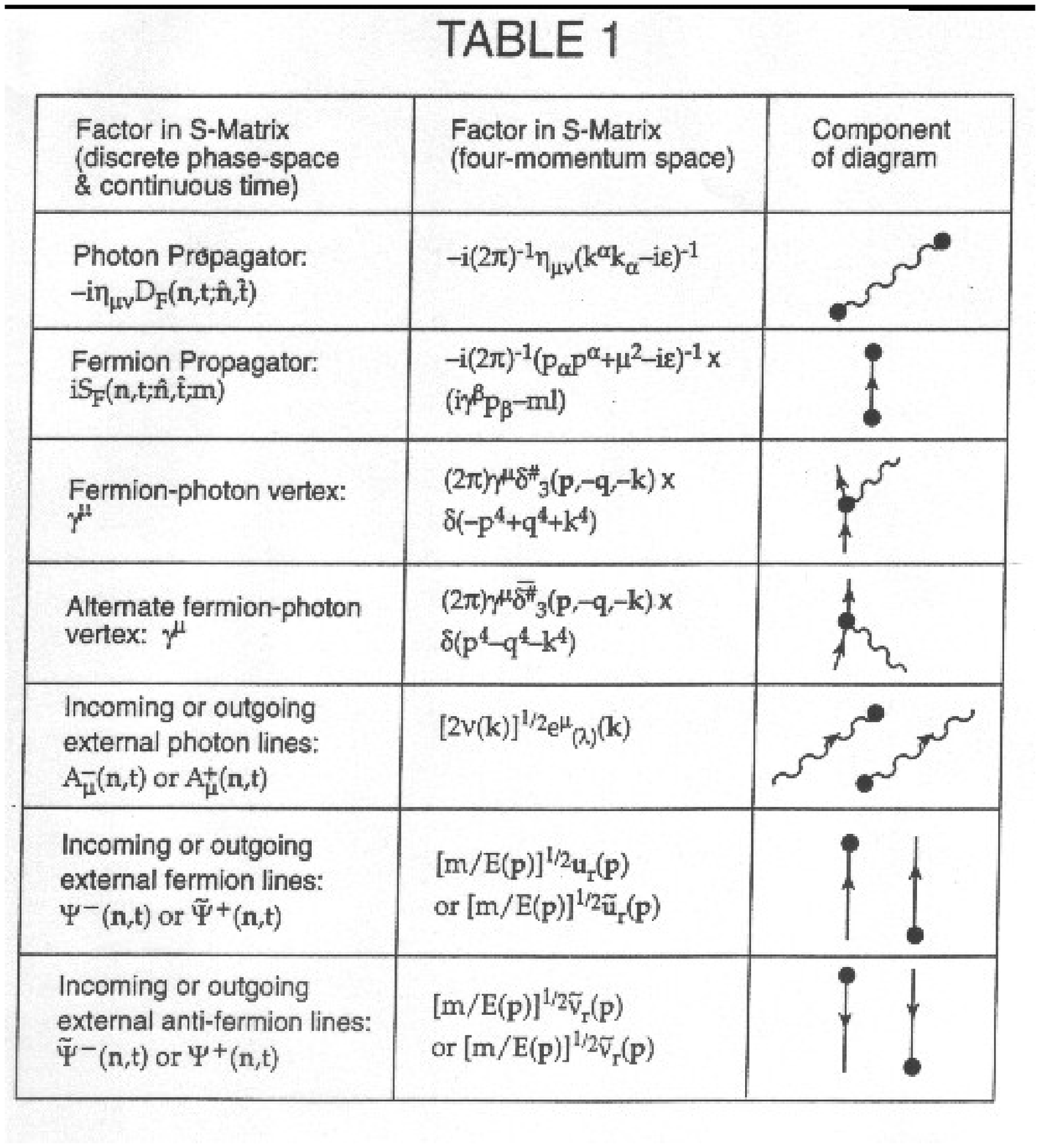}
\end{center}
{Table-1:{\small The second order photon self-energy graph in the momontum
space.}}
\end{figure}

\qquad The Feynman rules in the
four-momentum space (instead of the rules in the discrete phase space and
continuous time) will be used in the sequel for the sake of simplicity. 
\emph{Integration of all} the internal (or virtual) bosons and fermions (or
anti-fermions) must be performed over $\mathbf{R}^{4}$. (The fourth
component of the momentum is integrated over the real axis because of the
addition of $-i\epsilon $ in the propagators.) The limit $\epsilon
\rightarrow 0_{+}$ is taken after all the internal integrations are
performed.

\qquad However, the actual construction of a matrix element $<f|S_{j}|i>$ in
(\ref{eq:fourtwo}) from Table-I is somewhat incomplete until we determine
the correct numerical factor which multiplies the ordered product of
operators with the same $j$ and the same physical process. It can be deduced
that the multiplicative factor must be 
\begin{equation}
\kappa _{j}:=(-1)^{l}\hbox{sgn}(\sigma )(-q|e|)^{j}.  \label{eq:fourten}
\end{equation}
Here, $l$ is the number of closed loops, $\sigma $ is the permutation of the
final fermions, and $q=-1$ for electrons, $\pm 1/3$ for quarks etc. The
integer $j$ is the number of vertices. In case, $\delta _{3}^{\#}(\mathbf{p}%
,-\mathbf{k},-\mathbf{q})$ in Table-I is replaced by $(2\pi )^{3}\delta ^{3}(%
\mathbf{p}-\mathbf{k}-\mathbf{q})$, we obtain exactly the same Feynman rules
which emerge out of the usual theory. However, we shall prove in the
Appendix that $\delta _{3}^{\#}(\mathbf{p},-\mathbf{k},-\mathbf{q})\neq
(2\pi )^{3}\delta ^{3}(\mathbf{p}-\mathbf{k}-\mathbf{q})$. Feynman rules in
Table-I are all \emph{manifestly} relativistic except possibly the vertex
term.

\qquad According to the Feynman rules in Table-I, energy is precisely
conserved at each vertex. Therefore, we can introduce a physically
meaningful matrix $M^{\#}$ by: 
\begin{eqnarray}
\langle f|M^{\#}|i\rangle &:&=\langle f|\sum_{j=1}^{\infty
}M_{j}^{\#}|i\rangle ,  \nonumber \\
\langle f|S_{j}|i\rangle &=&:i(2\pi )\delta (E_{(f)}-E_{(i)})\langle
f|M_{j}^{\#}|i\rangle ,  \nonumber \\
\langle f|S-I|i\rangle &=&i(2\pi )\delta (E_{(f)}-E_{(i)})\langle
f|M^{\#}|i\rangle ,  \label{eq:foureleven}
\end{eqnarray}
where $E_{(i)}$ and $E_{(f)}$ are the initial and final energies
respectively.

\qquad The transition probability from the initial state $|i\rangle$ into
the final state $|f\rangle$ \emph{per unit time} is provided by: 
\begin{equation}
\omega_{(f)(i)}:=(2\pi)^{2}\delta(E_{(f)}-E_{(i)}) |\langle
f|M^{\#}|i\rangle|^{2}.  \label{eq:fourtwelve}
\end{equation}

\section{The second order self-energies of electron and photon}

\qquad For the analysis of the second order self-energy of an electron we
start with the equation (\ref{eq:fournine}). Moreover, utilizing (\ref
{eq:foureleven}) we can write 
\begin{equation}
\langle f|S_{2}|i\rangle =i(2\pi )\delta (\hat{E}-E)\langle
f|M_{2}^{\#}|i\rangle  \nonumber
\end{equation}
\begin{eqnarray}
&=&:-2e^{2}m\left[ E(\hat{\mathbf{p}})E(\mathbf{p})\right] ^{-1/2}\mathbf{%
\tilde{u}}_{s}(\hat{\mathbf{p}})\sum_{(2)}^{\#}(p,\hat{p})\mathbf{u}_{r}(%
\mathbf{p}),  \nonumber \\
-2\sum_{(2)}^{\#}(p,\hat{p}) &:&=\lim_{\epsilon \rightarrow 0_{+}}\int_{%
\mathbf{R}^{4}}\int_{\mathbf{R}^{4}}d^{4}q\,d^{4}k\,\overline{\delta
_{3}^{\#}(\mathbf{\hat{p},-q,-k})}\delta (\hat{E}-q^{4}-k^{4})  \nonumber \\
&&\times \delta _{3}^{\#}(\mathbf{p,-q,-k})\delta (-E+q^{4}+k^{4})\gamma
^{\mu }(i\gamma ^{\rho }q_{\rho }-mI)  \nonumber \\
&&\times (q^{\alpha }q_{\alpha }+m^{2}-i\epsilon )^{-1}(k^{\beta }k_{\beta
}-i\epsilon )^{-1}\gamma _{\mu }.  \label{eq:fiveone}
\end{eqnarray}

\qquad Using the properties of the Dirac matrices and the asymptotic
approximation $d_{3}^{\#}(\mathbf{p,q,k})$ in equation (A.18), we obtain
from (\ref{eq:fiveone}) that 
\begin{eqnarray}
\sum_{(2)}^{\#}(p,\hat{p}) &\simeq &\lim_{\epsilon \rightarrow 0_{+}}\int_{%
\mathbf{R}^{4}}\int_{\mathbf{R}^{4}}d^{4}q\,d^{4}k\,\overline{d_{3}^{\#}(%
\mathbf{\hat{p},-q,-k})}\delta (\hat{E}-q^{4}-k^{4})d_{3}^{\#}(\mathbf{%
p,-q,-k})  \nonumber \\
&&\delta (-E+q^{4}+k^{4})(i\gamma ^{\rho }q_{\rho }+2mI)(q^{\alpha
}q_{\alpha }+m^{2}-i\epsilon )^{-1}  \nonumber \\
&&\times (k^{\beta }k_{\beta }-i\epsilon )^{-1}.  \label{eq:fivetwo}
\end{eqnarray}
The last two factors in the above integrand can be combined by the parameter 
$x$ into 
\begin{equation}
(q^{\alpha }q_{\alpha }+m^{2}-i\epsilon )^{-1}(k^{\beta }k_{\beta
}-i\epsilon )^{-1}=\int_{0}^{1}dx\,\left[ x(q^{\alpha }q_{\alpha
}+m^{2})+(1-x)k^{\beta }k_{\beta }-i\epsilon \right] ^{-2}.
\label{eq:fivethree}
\end{equation}
Moreover, we make the simplifying assumptions 
\begin{equation}
\hat{\mathbf{p}}=\mathbf{p}=\mathbf{0}.  \label{eq:fivefour}
\end{equation}
By (\ref{eq:fivetwo} - \ref{eq:fivefour}) and (A.15) we obtain \newpage 
\[
{\sum }_{(2)}^{\#}(\mathbf{0},m;\mathbf{0},m)=\lim_{\epsilon \rightarrow
0_{+}}\int_{\mathbf{R}^{4}}\int_{\mathbf{R}^{4}}\int_{0}^{1}d^{4}q\,d^{4}k%
\,dx|d_{3}^{\#}(\mathbf{0,-q,-k})|^{2} 
\]
\[
\left[ \delta (m-q^{4}-k^{4})\right] ^{2}(i\gamma ^{\rho }q_{\rho
}+2mI)\left[ x(q^{\alpha }q_{\alpha }+m^{2})+(1-x)k^{\beta }k_{\beta
}-i\epsilon \right] ^{-1} 
\]
\[
=(4)^{-1}(4\pi )^{-6}\lim_{\epsilon \rightarrow
0_{+}}\int_{R^{4}}\int_{R^{4}}\int_{0}^{1}d^{4}k\,d^{4}q\,dx\left[ \delta
(m-q^{4}-k^{4})\right] ^{2}\prod_{a=1}^{3}\left( 4^{3}\{|k_{a}+q_{a}|\right. 
\]
\[
\left[ \delta (k_{a}+q_{a})\right] ^{2}+|k_{a}-q_{a}|\left[ \delta
(k_{a}-q_{a})\right] ^{2}+2|k_{a}^{2}-q_{a}^{2}|^{1/2}\delta
(k_{a}+q_{a})\delta (k_{a}-q_{a})\} 
\]
\begin{eqnarray}
&&+2(4)^{2}\{\delta (k_{a}+q_{a})+\delta
(k_{a}-q_{a})+|(k_{a}+q_{a})/(k_{a}-q_{a})|^{1/2}\delta (k_{a}+q_{a}) 
\nonumber \\
&&+|(k_{a}-q_{a})/(k_{a}+q_{a})|^{1/2}\delta (k_{a}-q_{a})\}  \nonumber \\
&&+\left. 4\left[
|k_{a}+q_{a}|^{-1}+|k_{a}-q_{a}|^{-1}+2|k_{a}^{2}-q_{a}^{2}|^{-1/2}\right]
\right) (i\gamma ^{\rho }q_{\rho }+2mI)  \nonumber \\
&&\left[ x(q^{\alpha }q_{\alpha }+m^{2})+(1-x)k^{\beta }k_{\beta }-i\epsilon
\right] ^{-2}.  \label{eq:fivefive}
\end{eqnarray}

\qquad To analyze the above integral we employ some mathematical devices.
(i) We put $\sqrt{|k|}\delta (k)=|k|\left[ \delta (k)\right] ^{2}\equiv 0$, $%
\sqrt{|(k+q)/(k-q)|}\delta (k+q)=0$ for $k\neq -q$. (ii) Whenever there is a
factor in the integrand which is \emph{odd} with respect to some integration
variable, we \emph{drop} that term. With such tricks, the RHS of (\ref
{eq:fivefive}) can be reduced to the form 
\begin{equation}
4{\sum }_{(2)}^{\#}(\mathbf{0},m;\mathbf{0},m)=(4)^{3}(4\pi )^{-6}\delta
(0)\left[ 4^{6}A+4^{4}(B_{1}+B_{2}+B_{3})+4^{2}(C_{1}+C_{2}+C_{3})+D\right] ,
\nonumber
\end{equation}
\begin{equation}
A:=\lim_{\epsilon \rightarrow 0_{+}}\int_{R^{4}}d^{4}k\int_{0}^{1}dx\left[
-i\gamma ^{4}(m-k^{4})+2mI\right] \left[
k^{b}k_{b}-(k^{4}-mx)^{2}+m^{2}x^{2}-i\epsilon \right] ^{-2},  \nonumber
\end{equation}
\begin{eqnarray}
B_{1} &:&=\lim_{\epsilon \rightarrow
0_{+}}\int_{R}dq_{1}\int_{R^{4}}d^{4}k\int_{0}^{1}dx\left[ -i\gamma
^{4}(m-k^{4})+2mI\right] \left[ k^{b}k_{b}-(k^{4}-mx)^{2}+\right.  \nonumber
\\
&&\left. m^{2}x^{2}+x(q_{1}^{2}-k_{1}^{2})-i\epsilon \right] ^{-2}\left[
|k_{1}+q_{1}|+|k_{1}-q_{1}|+2|k_{1}^{2}-q_{1}^{2}|^{-1/2}\right]  \nonumber
\end{eqnarray}
\begin{eqnarray}
C_{1} &:&=\lim_{\epsilon \rightarrow
0_{+}}\int_{R^{2}}dq_{1}\,dq_{2}\int_{R^{4}}d^{4}k\int_{0}^{1}dx\left[
-i\gamma ^{4}(m-k^{4})+2mI\right] \left[ k^{b}k_{b}-(k^{4}-mx)^{2}+\right. 
\nonumber \\
&&\left. m^{2}x^{2}+x(q_{1}^{2}+q_{2}^{2}-k_{1}^{2}-k_{2}^{2})-i\epsilon
\right] ^{-2}\left[
|k_{2}+q_{2}|^{-1}+|k_{2}-q_{2}|^{-1}+2|k_{2}^{2}-q_{2}^{2}|^{-1/2}\right] ,
\nonumber
\end{eqnarray}
\begin{eqnarray}
D &:&=\lim_{\epsilon \rightarrow 0_{+}}\int_{R^{2}}d^{3}\mathbf{q}%
\,\int_{R^{4}}d^{4}k\int_{0}^{1}dx\left[ -i\gamma ^{4}(m-k^{4})+2mI\right]
\left[ k^{b}k_{b}-(k^{4}-mx)^{2}+m^{2}x^{2}\right.  \nonumber \\
&&\left. +x(q^{c}q_{c}-k^{c}k_{c})-i\epsilon \right] ^{-2}\left[
\prod_{a=1}^{3}\left(
|k_{a}+q_{a}|^{-1}+|k_{a}-q_{a}|^{-1}+2|k_{a}^{2}-q_{a}^{2}|^{-1/2}\right)
\right] .  \label{eq:fivesix}
\end{eqnarray}

\qquad In the above expression, a mathematically unacceptable symbol, $%
\delta (0)$ occurs. We can explain it by physical arguments of energy
conservation $\delta (\hat{E} - E) = \delta (0)$ in (\ref{eq:fivetwo}) and (%
\ref{eq:foureleven}).

\qquad We note that $A, B_{1}, C_{1}, D$ are all ``logarithmically
divergent" integrals. Therefore, symbolically we can write: 
\begin{equation}
{\sum}_{(2)}^{\#}(\mathbf{0},m;\mathbf{0},m)=m \delta (0) \left[ 
\hbox{log. div.
terms}\right].  \label{eq:fiveseven}
\end{equation}
From the $S_{2}$- matrix element of the usual quantum field theory in the
space-time continuum, we obtain 
\begin{eqnarray}
{\sum}_{(2)}(\mathbf{0},m;\mathbf{0},m) &=& m \left[\delta (0)\right]^{4}
\int_{\mathbf{R}^4}d^{4}k \int_{0}^{1} dx (k^{\beta}k_{\beta} + m^{2}x^{2}
)^{-2} \left[2I -i \gamma^{4}(1-x)\right]  \nonumber \\
&=& m\left[\delta (0)\right]^{4} \left[ \hbox{log. div. terms} \right].
\label{eq:fiveeight}
\end{eqnarray}
Comparing (\ref{eq:fiveseven}) and (\ref{eq:fiveeight}) we conclude that the 
$S_2$ matrix element for the fermionic self-energy case in the discrete
phase space and continuous time does converge!

\qquad We shall now work out the second order photon self-energy (see fig.2.
Below.) 

\begin{figure}[ht]
\begin{center}
\includegraphics[bb=200 244 329 604, scale=0.4, clip, keepaspectratio=true]{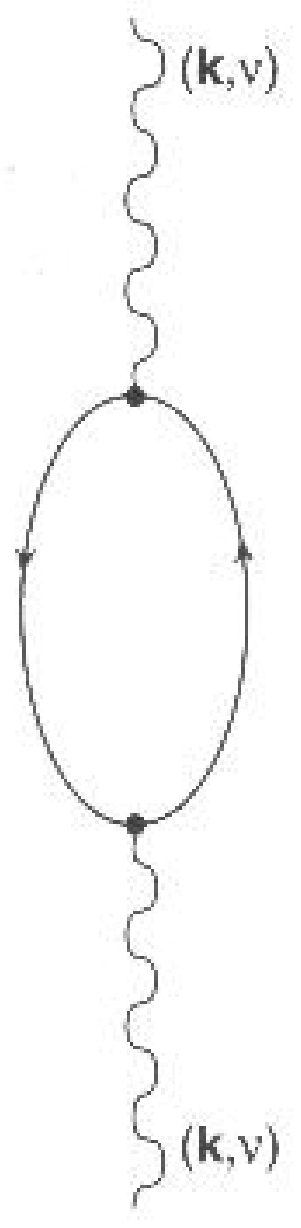}
\caption{{\small The second order photon self-energy graph in the momontum
space.}}
\label{fig:two}
\end{center}
\end{figure}

\qquad This self energy is given by the Feynman rules in Table-I as
proportional to 
\begin{eqnarray}
\Pi ^{\#\mu \nu }(\mathbf{k}) &:&=\lim_{\epsilon \rightarrow 0_{+}}\{%
\hbox{Tr}\int_{\mathbf{R}^{4}}\int_{\mathbf{R}^{4}}d^{4}p\,d^{4}\hat{p}%
\gamma ^{\mu }\left[ \frac{i\gamma ^{\alpha }p_{\alpha }-mI}{p^{\beta
}p_{\beta }+m^{2}-i\epsilon }\right]  \nonumber \\
&&\times \left[ \frac{i\gamma ^{\rho }\hat{p}_{\rho }-mI}{\hat{p}^{\sigma }%
\hat{p}_{\sigma }+m^{2}-i\epsilon }\right] \gamma ^{\nu }\left[ \delta
(-p^{4}+\widehat{p}^{4}+\nu \right] ^{2}  \nonumber \\
&&\times \prod_{c=1}^{3}|\delta ^{\#}(-p_{c},\hat{p}_{c},k_{c})|^{2}.
\label{eq:fivenine}
\end{eqnarray}
\qquad We use the asymptotic approximation (A-I.21) to evaluate $\Pi ^{\#\mu
\nu }(\mathbf{k})$ above. It is given by (dropping $\epsilon $) 
\begin{eqnarray}
\Pi ^{\#\mu \nu }(\mathbf{k}) &\approx &-(4\pi )^{-6}\delta (0)\hbox{Tr}%
\{\gamma ^{\mu }\left[ 8\hat{A}+2\left( \hat{B}_{1}+\hat{B}_{2}+\hat{B}%
_{3}\right) \right.  \nonumber \\
&+&\left. (2)^{-1}\left( \hat{C}_{1}+\hat{C}_{2}+\hat{C}_{3}\right) +(8)^{-1}%
\hat{D}\right] \gamma ^{\nu }\},
\end{eqnarray}
\begin{eqnarray}
\hat{A}:= &&\int_{\mathbf{R}^{4}}d^{4}p\left[ i\gamma ^{\alpha }p_{\alpha
}-mI\right] \left[ i\gamma ^{4}(p^{4}-\nu )+mI\right] \left[ p^{\beta
}p_{\beta }+m^{2}\right] ^{-1}  \nonumber \\
&\times &\{\left[
(p_{1}+k_{1})^{2}+(p_{2}+k_{2})^{2}+(p_{3}+k_{3})^{2}-(p_{4}-\nu
)^{2}+m^{2}\right] ^{-1}  \nonumber \\
&+&\left[ (p_{1}+k_{1})^{2}+(p_{2}+k_{2})^{2}+(p_{3}-k_{3})^{2}-(p_{4}-\nu
)^{2}+m^{2}\right] ^{-1}  \nonumber \\
&+&\left[ (p_{1}+k_{1})^{2}+(p_{2}-k_{2})^{2}+(p_{3}+k_{3})^{2}-(p_{4}-\nu
)^{2}+m^{2}\right] ^{-1}  \nonumber \\
&+&\left[ (p_{1}+k_{1})^{2}+(p_{2}-k_{2})^{2}+(p_{3}-k_{3})^{2}-(p_{4}-\nu
)^{2}+m^{2}\right] ^{-1}  \nonumber \\
&+&\left[ (p_{1}-k_{1})^{2}+(p_{2}+k_{2})^{2}+(p_{3}+k_{3})^{2}-(p_{4}-\nu
)^{2}+m^{2}\right] ^{-1}  \nonumber \\
&+&\left[ (p_{1}-k_{1})^{2}+(p_{2}+k_{2})^{2}+(p_{3}-k_{3})^{2}-(p_{4}-\nu
)^{2}+m^{2}\right] ^{-1}  \nonumber \\
&+&\left[ (p_{1}-k_{1})^{2}+(p_{2}-k_{2})^{2}+(p_{3}+k_{3})^{2}-(p_{4}-\nu
)^{2}+m^{2}\right] ^{-1}  \nonumber \\
&+&\left[ (p_{1}-k_{1})^{2}+(p_{2}-k_{2})^{2}+(p_{3}-k_{3})^{2}-(p_{4}-\nu
)^{2}+m^{2}\right] ^{-1}\},  \nonumber
\end{eqnarray}
\begin{eqnarray}
\hat{B}_{1}:= &&\int_{\mathbf{R}^{4}}d^{4}p\int_{\mathbf{R}}d\hat{p}%
_{1}\left[ \frac{i\gamma ^{\alpha }p_{\alpha }-mI}{p^{\beta }p_{\beta }+m^{2}%
}\right] \left[ i\gamma ^{4}(p^{4}-\nu )+mI\right]  \nonumber \\
&\times &\{\left[ \left[ \hat{p}%
_{1}^{2}+(p_{2}+k_{2})^{2}+(p_{3}+k_{3})^{2}-(p^{4}-\nu )^{2}+m^{2}\right]
^{-1}\right.  \nonumber \\
&+&\left[ \hat{p}_{1}^{2}+(p_{2}+k_{2})^{2}+(p_{3}-k_{3})^{2}-(p^{4}-\nu
)^{2}+m^{2}\right] ^{-1}  \nonumber \\
&+&\left[ \hat{p}_{1}^{2}+(p_{2}-k_{2})^{2}+(p_{3}+k_{3})^{2}-(p^{4}-\nu
)^{2}+m^{2}\right] ^{-1}  \nonumber \\
&+&\left. \left[ \hat{p}_{1}^{2}+(p_{2}-k_{2})^{2}+(p_{3}-k_{3})^{2}-(p^{4}-%
\nu )^{2}+m^{2}\right] ^{-1}\right]  \nonumber \\
&\times &\left[ |p_{1}-\hat{p}_{1}-k_{1}|^{-1}+|p_{1}+\hat{p}%
_{1}+k_{1}|^{-1}+|p_{1}-\hat{p}_{1}+k_{1}|^{-1}+|p_{1}+\hat{p}%
_{1}-k_{1}|^{-1}\right.  \nonumber \\
&+&(|p_{1}-\hat{p}_{1}-k_{1})(p_{1}+\hat{p}_{1}+k_{1})|^{-1/2}\left( 1+%
\hbox{sgn}(p_{1}-\hat{p}_{1}-k_{1})\hbox{sgn}(p_{1}+\hat{p}_{1}+k_{1})\right)
\nonumber \\
&+&(|p_{1}-\hat{p}_{1}-k_{1})(p_{1}-\hat{p}_{1}+k_{1})|^{-1/2}\left( 1+%
\hbox{sgn}(p_{1}-\widehat{p}_{1}-k_{1})\hbox{sgn}(p_{1}-\hat{p}%
_{1}+k_{1})\right)  \nonumber \\
&+&(|p_{1}-\hat{p}_{1}-k_{1})(p_{1}+\hat{p}_{1}-k_{1})|^{-1/2}\left( 1+%
\hbox{sgn}(p_{1}-\hat{p}_{1}-k_{1})\hbox{sgn}(p_{1}+\hat{p}_{1}-k_{1})\right)
\nonumber \\
&+&(|p_{1}+\hat{p}_{1}+k_{1})(p_{1}-\hat{p}_{1}+k_{1})|^{-1/2}\left( 1+%
\hbox{sgn}(p_{1}+\hat{p}_{1}+k_{1})\hbox{sgn}(p_{1}-\hat{p}_{1}+k_{1})\right)
\nonumber \\
&+&(|p_{1}+\hat{p}_{1}+k_{1})(p_{1}+\hat{p}_{1}-k_{1})|^{-1/2}\left( 1+%
\hbox{sgn}(p_{1}+\hat{p}_{1}+k_{1})\hbox{sgn}(p_{1}+\hat{p}_{1}-k_{1})\right)
\nonumber \\
&+&\left. (|p_{1}-\hat{p}_{1}+k_{1})(p_{1}+\hat{p}_{1}-k_{1})|^{-1/2}\left(
1+\hbox{sgn}(p_{1}-\widehat{p}_{1}+k_{1})\hbox{sgn}(p_{1}+\hat{p}%
_{1}-k_{1})\right) \right] \}  \nonumber
\end{eqnarray}
\begin{eqnarray}
\hat{C}:= &&\int_{\mathbf{R}^{4}}d^{4}p\int_{\mathbf{R}^{2}}d\hat{p}_{1}\,d%
\hat{p}_{2}\left[ i\gamma ^{\alpha }p_{\alpha }-mI\right] \left[ i\gamma
^{4}(p^{4}-\nu )+mI\right] \left[ p^{\beta }p_{\beta }+m^{2}\right] ^{-1} 
\nonumber \\
&\times &\{\left[ \hat{p}_{1}^{2}+\hat{p}_{2}^{2}+(p_{3}+k_{3})^{2}-(p^{4}-%
\nu )^{2}+m^{2}\right] ^{-1}+\left[ \hat{p}_{1}^{2}+\hat{p}%
_{2}^{2}+(p_{3}-k_{3})^{2}-(p^{4}-\nu )^{2}+m^{2}\right] ^{-1}  \nonumber \\
&\times &\{\prod_{A=1}^{2}\left[ |p_{A}+\hat{p}_{A}+k_{A}|^{-1}+|p_{A}-\hat{p%
}_{A}-k_{A}|^{-1}+|p_{A}+\hat{p}_{A}-k_{A}|^{-1}+|p_{A}-\hat{p}%
_{A}-k_{A}|^{-1}\right.  \nonumber \\
&+&|(p_{A}+\hat{p}_{A}+k_{A})(p_{A}-\hat{p}_{A}-k_{A})|^{-1/2}\left( 1+%
\hbox{sgn}(p_{A}+\hat{p}_{A}+k_{A})\hbox{sgn}(p_{A}-\hat{p}_{A}-k_{A})\right)
\nonumber \\
&+&|(p_{A}+\hat{p}_{A}+k_{A})(p_{A}+\hat{p}_{A}-k_{A})|^{-1/2}\left( 1+%
\hbox{sgn}(p_{A}+\hat{p}_{A}+k_{A})\hbox{sgn}(p_{A}+\hat{p}_{A}-k_{A})\right)
\nonumber \\
&+&|(p_{A}+\hat{p}_{A}+k_{A})(p_{A}-\hat{p}_{A}+k_{A})|^{-1/2}\left( 1+%
\hbox{sgn}(p_{A}+\hat{p}_{A}+k_{A})\hbox{sgn}(p_{A}-\hat{p}_{A}+k_{A})\right)
\nonumber \\
&+&|(p_{A}-\hat{p}_{A}-k_{A})(p_{A}+\hat{p}_{A}-k_{A})|^{-1/2}\left( 1-%
\hbox{sgn}(p_{A}-\hat{p}_{A}-k_{A})\hbox{sgn}(p_{A}+\hat{p}_{A}-k_{A})\right)
\nonumber \\
&+&|(p_{A}-\hat{p}_{A}-k_{A})(p_{A}-\hat{p}_{A}+k_{A})|^{-1/2}\left( 1-%
\hbox{sgn}(p_{A}-\hat{p}_{A}-k_{A})\hbox{sgn}(p_{A}-\hat{p}_{A}+k_{A})\right)
\nonumber \\
&+&\left. |(p_{A}+\hat{p}_{A}-k_{A})(p_{A}-\hat{p}_{A}+k_{A})|^{-1/2}\left(
1+\hbox{sgn}(p_{A}+\hat{p}_{A}-k_{A})\hbox{sgn}(p_{A}-\hat{p}%
_{A}+k_{A})\right) \right] \}  \nonumber
\end{eqnarray}
\begin{eqnarray}
\hat{D}:= &&\int_{\mathbf{R}^{4}}d^{4}p\,\int_{\mathbf{R}^{3}}d^{3}\widehat{%
\mathbf{p}}\left[ i\gamma ^{\alpha }p_{\alpha }-mI\right] \left[ i\gamma
^{4}(p^{4}-\nu )+mI\right]  \nonumber  \label{eq:fiveonezero} \\
&\times &\left[ p^{\beta }p_{\beta }+m^{2}\right] ^{-1}\left[ \hat{p}_{b}%
\hat{p}_{b}-(p^{4}-\nu )^{2}+m^{2}\right] ^{-1}  \nonumber \\
&\times &\{\prod_{c=1}^{3}\left[ |p_{c}-\hat{p}_{c}-k_{c}|^{-1}+|p_{c}+\hat{p%
}_{c}+k_{c}|^{-1}+|p_{c}-\hat{p}_{c}+k_{c}|^{-1}+|p_{c}+\hat{p}%
_{c}-k_{c}|^{-1}\right.  \nonumber \\
&+&|(p_{c}-\hat{p}_{c}-k_{c})(p_{c}-\hat{p}_{c}-k_{c})|^{-1/2}\left( 1+%
\hbox{sgn}(p_{c}-\hat{p}_{c}-k_{c})\,\hbox{sgn}(p_{c}+\hat{p}%
_{c}+k_{c})\right)  \nonumber \\
&+&|(p_{c}-\hat{p}_{c}-k_{c})(p_{c}-\hat{p}_{c}+k_{c})|^{-1/2}\left( 1-%
\hbox{sgn}(p_{c}-\hat{p}_{c}-k_{c})\,\hbox{sgn}(p_{c}-\hat{p}%
_{c}+k_{c})\right)  \nonumber \\
&+&|(p_{c}-\hat{p}_{c}-k_{c})(p_{c}+\hat{p}_{c}-k_{c})|^{-1/2}\left( 1-%
\hbox{sgn}(p_{c}-\hat{p}_{c}-k_{c})\,\hbox{sgn}(p_{c}+\hat{p}%
_{c}-k_{c})\right)  \nonumber \\
&+&|(p_{c}+\hat{p}_{c}+k_{c})(p_{c}-\hat{p}_{c}+k_{c})|^{-1/2}\left( 1-%
\hbox{sgn}(p_{c}+\hat{p}_{c}+k_{c})\,\hbox{sgn}(p_{c}-\hat{p}%
_{c}+k_{c})\right)  \nonumber \\
&+&|(p_{c}+\hat{p}_{c}+k_{c})(p_{c}+\hat{p}_{c}-k_{c})|^{-1/2}\left( 1-%
\hbox{sgn}(p_{c}+\hat{p}_{c}+k_{c})\,\hbox{sgn}(p_{c}+\hat{p}%
_{c}-k_{c})\right)  \nonumber \\
&+&\left. |(p_{c}-\hat{p}_{c}+k_{c})(p_{c}+\hat{p}_{c}-k_{c})|^{-1/2}\left(
1+\hbox{sgn}(p_{c}-\hat{p}_{c}+k_{c})\,\hbox{sgn}(p_{c}+\hat{p}%
_{c}-k_{c})\right) \right] \}  \nonumber \\
&&
\end{eqnarray}

\qquad The analysis of the integrands in $\hat{A},\,\hat{B}_{a},\,\hat{C}%
_{a},\,\hat{D}$ imply that 
\begin{equation}
\Pi ^{\#\mu \nu }=\delta (0)\left[ \hbox{quadratically divergent terms}%
\right] .  \label{eq:fiveoneone}
\end{equation}
However, the expression from the usual theory in continuous space-time
indicates that 
\begin{equation}
\Pi ^{\mu \nu }=\left[ \delta (0)\right] ^{4}\left[ 
\hbox{quadratically
divergent terms}\right] .  \label{eq:fiveonetwo}
\end{equation}
Comparing (\ref{eq:fiveoneone}) and (\ref{eq:fiveonetwo}), we conclude that
the second order photon self-energy integral converges in the discrete phase
space and continuous time formulation.

\section{Convergence of the S-matrix element $\left\langle f\left|
S_{j}\right| i\right\rangle $}

\qquad We start this section with some simple examples. Let us consider the
improper Riemann integral $\int_{-\infty }^{\infty }k\,dk$. This integral
does \emph{not }converge. However, the Cauchy principal value C.P.V. $%
\int_{-\infty }^{\infty }k\,dk$ ``converges'' to zero. Now consider another
improper integral $\int_{-\infty }^{\infty }k^{2}\,dk$. It diverges
(strongly) and its Cauchy principal value does \emph{not} converge either.
Now let us investigate the infinite series $%
1-x+2x^{2}-(3!)x^{3}+(4!)x^{4}-...$ . It is a divergent series for $x\neq 0$%
. However, it is Borel-summable to the value $\int_{0}^{\infty
}e^{-t}[1+xt]^{-1}dt$ (which converges). Consider another (complex) infinite
series $\sum\limits_{n=0}^{\infty }\exp [i2^{n}x]$. It is strongly divergent
for all $x\in \mathbf{R}$ and cannot be ``summed'' in any sense.

In this section, our primary goal is to investigate the convergence of the $%
j $-$th$ order S-matrix element $\left\langle f\left| S_{j}\right|
i\right\rangle $, where $|i>$ and $|f>$ are initial and final states
respectively. This matrix element is furnished by the equation (\ref
{eq:fourtwo}). We apply Wick's theorem inherent in (\ref{eq:threeseven}) and
consider only terms with a specific number of contractions. There exist many
terms even in that category. We pick a typical term $\left\langle f\left| 
\overline{S}_{j}\right| i\right\rangle $ in a particular category. Using the
Table-I, we can express the matrix element as the following infinite series
of integrals: 
\begin{eqnarray}
\left\langle f\left| \overline{S}_{j}\right| i\right\rangle &=&(const)\sum_{%
\mathbf{n}_{1}=\mathbf{0}}^{\infty (3)}\int_{\mathbf{R}}dt_{1}...\sum_{%
\mathbf{n}_{j}=\mathbf{0}}^{\infty (3)}\int_{\mathbf{R}}dt_{j}\{<f|\,N\,[%
\gamma ^{\mu _{1}}\psi (\mathbf{n}_{1},t_{1})A_{\mu _{1}}(\mathbf{n}%
_{1},t_{1})  \nonumber \\
&&\gamma ^{\mu _{2}}S_{F}(\mathbf{n}_{1},t_{1};\mathbf{n}_{2},t_{2};m)\eta
_{\mu _{2}\mu _{3}}D_{F}(\mathbf{n}_{2},t_{2};\mathbf{n}_{3},t_{3})S_{F}(%
\mathbf{n}_{2},t_{2};\mathbf{n}_{3},t_{3};m)\gamma ^{\mu _{3}}...  \nonumber
\\
&&S_{F}(\mathbf{n}_{j-1},t_{j-1};\mathbf{n}_{j},t_{j};m)\widetilde{\psi }%
\gamma ^{\mu _{j}}A_{\mu _{j}}(\mathbf{n}_{j},t_{j})]\}.  \label{eq:sixone}
\end{eqnarray}

The right hand side of the above equation contains multiple series and
multiple integrals. There exist three possibilities. It can be convergent,
or summable in some sense, or strongly divergent. Before embarking upon such
investigations, let us consider a similar problem for a much simpler series
of integrals: 
\[
\sum_{n=0}^{\infty }\left[ c_{n}\int_{-\infty }^{\infty }f_{n}(k)dk\right] . 
\]
A necessary criterion for convergence is that $\int_{-\infty }^{\infty
}f_{n}(k)dk$ converges (or converges as Cauchy principal value) for each $%
n\in \left\{ 0,1,2,...\right\} $ to $s_{n}$. If furthermore, $%
\sum_{0}^{\infty }(c_{n}s_{n})$ converges, then the original series of
integrals is said to converge. In case the series $\sum_{0}^{\infty
}(c_{n}s_{n})$ does \emph{not }converge, we should determine whether it is
summable or strongly divergent. (In the case of the right hand side of (\ref
{eq:sixone}), such analysis is extremely cumbersome.) In case the analysis
of $\sum_{0}^{\infty }(c_{n}s_{n})$ is very difficult, we can try the \emph{%
asymptotic} analysis. In this strategy, criteria for ``convergence'' of $%
\sum_{n=0}^{\infty }\left[ c_{n}\int_{-\infty }^{\infty }f_{n}(k)dk\right] $
is summarized by the following steps:

i) Prove that C.P.V.$\int_{-\infty }^{\infty }f_{n}(k)dk$ converges to $%
s_{n} $ for every $n\in \left\{ 0,1,2,...\right\} $.

ii) Express the series as $\sum_{n=0}^{N}\left[ c_{n}\int_{-\infty }^{\infty
}f_{n}(k)dk\right] +\sum_{n=N+1}^{\infty }\left[ c_{n}\int_{-\infty
}^{\infty }f_{n}(k)dk\right] $for a sufficiently large positive integer $N$.

iii) Prove that the series $\sum_{n=1}^{\infty }c_{N+n}f_{N+n}(k)$ converges
to $\sigma _{N}(k)$ .

iv) Check that $\lim\limits_{N\rightarrow \infty }\sigma _{N}(k)=0.$

v) Finally, prove that C.P.V.$\int_{-\infty }^{\infty }\sigma
_{N}(k)dk=\int_{-\infty }^{\infty }\left[ \sum_{n=N+1}^{\infty
}c_{n}f_{n}(k)\right] dk$ converges.

\medskip

We shall follow a similar five step strategy to investigate the convergence
of the matrix element $\left\langle f\left| \overline{S}_{j}\right|
i\right\rangle $ in (\ref{eq:sixone}). For the step i), we have to
investigate improper integrals representing propagators (See the Appendix-II
of the paper-II.) These are furnished by: 
\begin{eqnarray}
D_{(a)}(\mathbf{n},t;\hat{\mathbf{n}},\hat{t}) &=&(2\pi )^{-1}\int_{\mathbf{R%
}^{3}}\left[ \prod_{b=1}^{3}\xi _{n_{b}}(k_{b})\xi _{n_{b}}(k_{b})\right] 
\nonumber \\
&&\left[ \int_{C_{(a)}}\left( \eta ^{\alpha \beta }k_{\alpha }k_{\beta
}\right) ^{-1}\exp \left[ ik_{4}(t-\hat{t})\right] dk^{4}\right] d^{3}%
\mathbf{k},  \label{eq:sixtwoone}
\end{eqnarray}
\begin{eqnarray}
S_{(a)}(\mathbf{n},t;\hat{\mathbf{n}},\hat{t}) &=&(2\pi )^{-1}\int_{\mathbf{R%
}^{3}}\left[ \prod_{b=1}^{3}\xi _{n_{b}}(k_{b})\xi _{n_{b}}(k_{b})\right] 
\nonumber \\
&&\left[ \int_{C_{(a)}}\left( i\gamma ^{\mu }p_{\mu }+mI\right) ^{-1}\exp
\left[ ip_{4}(t-\hat{t})\right] dp^{4}\right] d^{3}\mathbf{p}.
\label{eq:sixtwotwo}
\end{eqnarray}
With the help of reference$^{10}$, we have integrated explicitly the right
hand side of (\ref{eq:sixtwoone}) in case of $D_{+}(\mathbf{n},t;\hat{%
\mathbf{n}},\hat{t})$. The result is the following: 
\begin{eqnarray}
&&D_{+}(\mathbf{n},t;\hat{\mathbf{n}},\hat{t})=-i\left[ 2(2\sqrt{\pi }%
)^{3}\right] ^{-1}\exp \left[ -(t-\hat{t})^{2}/8\right] \left[
\prod_{b=1}^{3}(i)^{n_{b}-\hat{n}_{b}}\sqrt{n_{b}!\hat{n}_{b}!}\right] 
\nonumber \\
&&\times \sum_{j_{1}=0}^{[n_{1}/2]}..\sum_{\hat{j}_{3}=0}^{[\hat{n}%
_{3}/2]}(-2)^{-(j_{1}+..+\hat{j}_{3})}\left[ 1+(-1)^{n_{1}+\hat{n}%
_{1}-2(j_{1}+\hat{j}_{1})}\right] \left[ 1+(-1)^{n_{2}+\hat{n}_{2}-2(j_{2}+%
\hat{j}_{2})}\right]  \nonumber \\
&&\times \left[ 1+(-1)^{n_{3}+\hat{n}_{3}-2(j_{3}+\hat{j}_{3})}\right]
\left[ j_{1}!..\hat{j}_{3}!(n_{1}-2j_{1})!..(\hat{n}_{3}-2\hat{j}%
_{3})!\right] ^{-1}  \nonumber \\
&&\times \Gamma \left[ 2^{-1}\left( n_{1}+\hat{n}_{1}-2(j_{1}+\hat{j}%
_{1})\right) +1\right] \Gamma \left[ 2^{-1}\left( n_{2}+\hat{n}_{2}-2(j_{2}+%
\hat{j}_{2})\right) +1\right]  \nonumber \\
&&\times \Gamma \left[ 2^{-1}\left( n_{3}+\hat{n}_{3}-2(j_{3}+\hat{j}%
_{3})\right) +1\right] \left( \Gamma \left[ 2^{-1}(n_{1}+...+\hat{n}%
_{3}-2(j_{1}+..+\hat{j}_{3}))+3\right] \right) ^{-1}  \nonumber \\
&&\times D_{-\left[ (n_{1}+...+\hat{n}_{3}-2(j_{1}+..+\hat{j}_{3}))+2\right]
}\left( i(t-\hat{t})/\sqrt{2}\right) ;  \label{eq:sixthreeone}
\end{eqnarray}
\begin{eqnarray}
\left[ \frac{n}{2}\right] &:&=n/2\;\;\;\;\;\mbox{for}\;\;n\;\;\mbox{even}, 
\nonumber \\
&&(n-1)/2\hspace{0.1in}\mbox{for\hspace{0.1in}}n\hspace{0.1in}\mbox{odd.}
\label{eq:sixthreetwo}
\end{eqnarray}

Here, $D_{-p}(iz)$ is the parabolic cylinder function$^{10}$.

The right hand side of (\ref{eq:sixthreeone}) is defined (or non-singular)
everywhere. In fact, in the coincident point, 
\begin{equation}
D_{+}\left( \mathbf{0},0;\mathbf{0},0\right) =-i\sqrt{{\pi }}=-D_{-}\left( 
\mathbf{0},0;\mathbf{0},0\right) .
\end{equation}
Thus, by equations (A.II.4B) and (A.II.5B) of the paper-II, $D_{F}\left( 
\mathbf{n},t;\hat{\mathbf{n}},\hat{t}\right) $ in (\ref{eq:sixtwoone}) \emph{%
is non-singular everywhere}. Similarly, the integral representing $%
S_{F}\left( \mathbf{n},t;\hat{\mathbf{n}},\hat{t};m\right) $ in (\ref
{eq:sixtwotwo}) converges by the equations 
\begin{eqnarray}
S_{F}\left( \mathbf{n},t;\hat{\mathbf{n}},\hat{t};m\right) &=&\left( \gamma
^{\mu }\Delta _{\mu }^{\#}-mI\right) \Delta _{F}\left( \mathbf{n},t;\hat{%
\mathbf{n}},\hat{t};m\right) ;  \nonumber \\
D_{F}\left( \mathbf{n},t;\hat{\mathbf{n}},\hat{t}\right) := &&\Delta
_{F}\left( \mathbf{n},t;\hat{\mathbf{n}},\hat{t};0\right) .
\end{eqnarray}
In fact, in the coincident points, 
\begin{eqnarray}
S_{+}\left( \mathbf{0},0;\mathbf{0},0;m\right) &=&2^{-1}\left[ -\gamma
^{4}+im^{3}\Psi (3/2,2;m^{2})I\right] ,  \nonumber \\
S_{-}\left( \mathbf{0},0;\mathbf{0},0;m\right) &=&2^{-1}\left[ -\gamma
^{4}-im^{3}\Psi (3/2,2;m^{2})I\right] .
\end{eqnarray}
Here, $\Psi (\alpha ,\beta ;z)$ is a degenerate hypergeometric function$%
^{10} $.

\qquad In the integrations for $D_{+}\left( \mathbf{n},t;\hat{\mathbf{n}},%
\hat{t}\right) $ in (\ref{eq:sixthreeone}) and $S_{\pm }\left( \mathbf{0},0;%
\mathbf{0},0;m\right) $ in (6.8) \emph{neither ultraviolet nor infrared
divergences are encountered}.

\qquad Since wave functions $A_{\mu }(\mathbf{n},t)$ and $\psi (\mathbf{n},t)
$ are defined everywhere, we can conclude that (initial) finite part of the
infinite series in (\ref{eq:sixone}), namely 
\begin{equation}
(const)\sum_{\mathbf{n}_{1}=\mathbf{0}}^{\mathbf{N}_{1}(3)}\int_{\mathbf{R}%
}dt_{1}...\sum_{\mathbf{n}_{j}=\mathbf{0}}^{\mathbf{N}_{j}(3)}\int_{\mathbf{R%
}}dt_{j}\{<f|..|i>\}  \label{eq:sixseven}
\end{equation}
is always \emph{non-singular}. We have to investigate the convergence of the
remaining tail-end: 
\begin{equation}
(const)\sum_{\mathbf{n}_{1}=\mathbf{N}_{\mathbf{1}}\mathbf{+1}}^{\infty
(3)}\int_{\mathbf{R}}dt_{1}...\sum_{\mathbf{n}_{j}=\mathbf{N}_{\mathbf{j}}%
\mathbf{+1}}^{\infty (3)}\int_{\mathbf{R}}dt_{j}\{<f|..|i>\};
\label{eq:sixeight}
\end{equation}
where $\mathbf{1}:=(1,1,1)$. For that purpose we take recourse to the
analogue of step iii) in the strategy. We have to prove that 
\begin{equation}
\sum_{\mathbf{n}_{1}=\mathbf{N}_{\mathbf{1}}\mathbf{+1}}^{\infty }\int_{%
\mathbf{R}}dt_{1}...\sum_{\mathbf{n}_{j}=\mathbf{N}_{\mathbf{j}}\mathbf{+1}%
}^{\infty (3)}\int_{\mathbf{R}^{4}}dt_{j}\int_{\mathbf{R}^{4}}d^{4}k_{1}...%
\int_{\mathbf{R}^{4}}d^{4}p_{1}..\{\mbox{Fourier Transform 
of}<f|..|i>\};
\label{eq:sixnine}
\end{equation}
converges. In this endeavour, we use Fourier transforms of all fields (as in
(4.5)). Thus we encounter at each vertex the modified distribution function: 
\begin{equation}
\delta _{3N}^{\#}(\mathbf{p},\mathbf{q},\mathbf{k}):=\sum_{n_{1}=N_{1}+1}^{%
\infty }\sum_{n_{2}=N_{2}+1}^{\infty }\sum_{n_{3}=N_{3}+1}^{\infty }\left[
\prod_{b=1}^{3}\xi _{n_{b}}(p_{b})\xi _{n_{b}}(q_{b})\xi
_{n_{b}}(k_{b})\right] .
\end{equation}
(Compare the above with the distribution function $\delta _{3}^{\#}(\mathbf{p%
},\mathbf{q},\mathbf{k})$ in the Appendix.) Let us first try a simpler
distribution function: 
\begin{equation}
\delta _{N}^{\#}(\mathbf{p},\mathbf{q},\mathbf{k}):=\sum_{n=N+1}^{\infty
}\xi _{n}(p)\xi _{n}(q)\xi _{n}(k)\;,\;\;(p,q,k)\in \mathbf{R}^{3}.
\label{eq:sixeleven}
\end{equation}
The exact sum in (\ref{eq:sixeleven}) is \emph{intractable} presently.
Therefore, we try the asymptotic approximation for a sufficiently large
positive $N$, as in the Appendix, by the expression 
\begin{eqnarray}
\delta _{N}^{\#}(p,q,k) &\simeq &d_{N}^{\#}(p,q,k):=\sum_{n=N+1}^{\infty
}\zeta _{n}(p)\zeta _{n}(q)\zeta _{n}(k)  \label{eq:sixtwelve} \\
&\simeq &(2\sqrt{\pi })^{-3}\sum_{n=N+1}^{\infty }n^{-3/4}\left[ e^{i2\sqrt{n%
}(p+q+k)}+e^{-i2\sqrt{n}(-p+q+k)}\right.   \nonumber \\
&&+\left. e^{-i2\sqrt{n}(p+q-k)}+e^{i2\sqrt{n}(-p+q-k)}\right] .  \nonumber
\end{eqnarray}
By the theorem A.1 of the Appendix, the infinite oscillatory series
converges as 
\begin{equation}
\sum_{n=N+1}^{\infty }n^{-3/4}e^{i2k\sqrt{n}}\simeq \int_{N}^{\infty
}y^{-3/4}e^{i2k\sqrt{y}}\,dy.  \label{eq:sixthirteen}
\end{equation}
The improper integral in (\ref{eq:sixthirteen}) converges$^{10}$ (except for 
$k\neq 0$) to 
\begin{eqnarray}
\int_{N}^{\infty }y^{-3/4}e^{i2k\sqrt{y}}\,dy &\simeq &-\{\sqrt{|k|}\delta
(k)+(1/4)|k|^{-1/2}  \nonumber \\
&&\times \left[ (1-i)\Gamma \left( -1/2,i2\sqrt{N}|k|\right) +(1+i)\Gamma
\left( -1/2,-i2\sqrt{N}|k|\right) \right]   \nonumber \\
&&-N^{-1/4}|k|^{-1}\sin (2\sqrt{N}|k|)  \nonumber \\
&&+i2^{-1}|k|^{-1/2}\mbox{sgn}(k)[(1+i)\Gamma (1/2,i2\sqrt{N}|k|)
\label{eq:sixfourteen} \\
&&+(1-i)\Gamma (1/2,-i2\sqrt{N}|k|)].  \nonumber
\end{eqnarray}
Here, $\Gamma (\alpha ,iz)$ is the incomplete gamma function. Using the
asymptotic representation of the incomplete gamma function$^{10}$ for very
large $\sqrt{N}|k|$, we derive from (\ref{eq:sixtwelve}) (\ref
{eq:sixthirteen}) (\ref{eq:sixfourteen}) that 
\begin{eqnarray}
\int_{N}^{\infty }y^{-3/4}e^{i2k\sqrt{y}}\,dy &=&N^{-1/4}\cos (2\sqrt{N}%
|k|)\delta (k)+i\mbox{sgn}(k)\exp [i2\sqrt{N}k]+O(1/N^{3/4}|k|^{2}) 
\nonumber \\
&\approx &(N^{-1/4}|k|^{-1}\left[ N^{-1/2}\delta 
(k)+i\mbox{sgn}(k)\right] .
\label{eq:sixfifteen}
\end{eqnarray}
Therefore, by the equations (\ref{eq:sixfourteen}) we conclude that 
\begin{eqnarray}
(2\sqrt{\pi })^{3}d_{N}^{\#}(p,q,k) &\simeq &N^{-1/4}\{|p+q+k|^{-1}\left[
N^{-1/2}\delta (p+q+k)+i\mbox{sgn}(p+q+k)\right]   \nonumber \\
&&+|q+k-p|^{-1}\left[ N^{-1/2}\delta (q+k-p)+i\mbox{sgn}(q+k-p)\right]  
\nonumber \\
&&+|q-k+p|^{-1}\left[ N^{-1/2}\delta (q-k+p)+i\mbox{sgn}(q-k+p)\right]  
\nonumber \\
&&+|q-k-p|^{-1}\left[ N^{-1/2}\delta (q-k-p)+i\mbox{sgn}(q-k-p)\right] 
\} 
\nonumber \\
&=&:\sigma _{N}(p,q,k).  \label{eq:sissixteen}
\end{eqnarray}
Thus, by equations (\ref{eq:sixtwelve}) and (\ref{eq:sixfifteen}) we obtain
the asymptotic approximation 
\begin{eqnarray}
(8\pi ^{3/2})^{3}\delta _{3N}^{\#}(\mathbf{p},\mathbf{q},\mathbf{k}) &\simeq
&(8\pi ^{3/2})^{3}d_{3N}^{\#}(\mathbf{p},\mathbf{q},\mathbf{k})  \nonumber \\
&:&=(8\pi ^{3/2})^{3}\prod_{b=1}^{3}d_{N_{b}}^{\#}(p_{b},q_{b},k_{b}) 
\nonumber \\
&\simeq &\prod_{b=1}^{3}\sigma _{N_{b}}(p_{b},q_{b},k_{b})  \nonumber \\
&=&\prod_{b=1}^{3}\{N_{b}^{-1/4}|p_{b}+q_{b}+k_{b}|^{-1}[N_{b}^{-1/2}\delta
(p_{b},q_{b},k_{b})+  \nonumber \\
&&i\mbox{sgn}(p_{b}+q_{b}+k_{b})]+..+..+..\}.  \label{eq:sixseventeen}
\end{eqnarray}
For completion of the step iii) in the strategy, we need to consider $j$
products of the triple series as in (6.12). These multiple series obviously
converge asymptotically to the products of terms as in \ref{eq:sixseventeen}.

As for the step iv) of the strategy, we notice that the analogous necessary
condition 
\begin{equation}
\lim_{(N_{1},N_{2},N_{3})\rightarrow (\infty ,\infty ,\infty )}\left[
\prod_{b=1}^{3}\sigma _{N_{b}}(p_{b},q_{b},k_{b})\right] =0
\label{eq:sixeighteen}
\end{equation}
is satisfied by (\ref{eq:sixseventeen}).

Now we have to investigate the most important step v) of the strategy.
Suppose that for the S-matrix element $\left\langle f\left| \overline{S}%
_{j}\right| i\right\rangle $ in (\ref{eq:sixone}) yielding the equation
(6.10), the corresponding Feynman graph involves the following
specifications:

$E_{F}$=The number of external electron-positron or fermion lines,

$E_{B}$=The number of external photon or boson lines,

$I_{F}$=The number of internal fermion lines,

$I_{B}$=The number of internal boson lines,

$j$=The number of corner or vertices.

For a trilinear interaction as in the equation (\ref{eq:fourone}), the
following equations$^{4}$ among various numbers hold: 
\begin{eqnarray}
I_{F} &=&j-\left( E_{F}/2\right) \geq 0,  \nonumber \\
I_{B} &=&\left( j-E_{B}\right) /2\geq 0,  \nonumber \\
j &=&I_{F}+\left( E_{F}/2\right) =2I_{B}+E_{B}\geq 0.  \label{eq:sixnineteen}
\end{eqnarray}
The typical S-matrix element in (6.10), with the help of the modified
Feynman rules in momentum space (Table-I), yields the asymptotic
approximation: 
\begin{eqnarray}
<f|\overline{S}_{j_{N}}|i> &\simeq &(const)\int_{\mathbf{R}%
^{4I_{B}}}d^{4}k_{(1)}..d^{4}k_{(I_{B})}\int_{\mathbf{R}%
^{4I_{F}}}d^{4}p_{(1)}..d^{4}p_{(I_{F})}  \nonumber \\
&&\times \left\{ \gamma ^{\rho }\left[ \gamma ^{\mu }p_{(1)\mu }+mI\right]
^{-1}..\left[ \gamma ^{\nu }p_{(I_{F})\nu }+mI\right] ^{-1}\gamma _{\rho
}\right.  \nonumber \\
&&\times \left[ k_{(1)}^{\alpha }k_{(1)\,\alpha }\right] ^{-1}..\left[
k_{(I_{B})}^{\beta }k_{(I_{B})\,\beta }\right] ^{-1}  \nonumber \\
&&\times d_{3N}^{\#}\left( \mathbf{p}_{(1)},-\mathbf{p}_{(2)},\mathbf{k}%
\right) \delta \left( p_{(1)_{4}}-p_{(2)_{4}}+k_{(1)_{4}}\right) .. 
\nonumber \\
&&\left. d_{3N}^{\#}\left( -\mathbf{p}_{(j-1)},\mathbf{p}_{(j)},\mathbf{k}%
_{(j)}\right) \delta \left( -p_{(j-1)_{4}},p_{(j)_{4}},k_{(j)_{4}}\right)
\right\} .  \label{eq:sixtwenty}
\end{eqnarray}
Integrating over all $p_{(l)_{4}}\equiv -E_{(l)}$ and $k_{(l)_{4}}\equiv
-\nu _{(l)}$ variables we get from (\ref{eq:sixtwenty}) 
\begin{eqnarray}
<f|\overline{S}_{j_{N}}|i> &=&(const.)\,\delta \left( \sum \left(
E_{(l)}+\nu _{(l)}\right) \right) \int_{\mathbf{R}^{3I_{B}}}d^{3}\mathbf{k}%
_{(1)}..d^{3}\mathbf{k}_{(I_{B})}\int_{\mathbf{R}^{3I_{F}}}d^{3}\mathbf{p}%
_{(1)}..d^{3}\mathbf{p}_{(I_{F})}  \nonumber \\
&&\times \int_{\mathbf{R}^{I_{B}+I_{F}-j+1}}%
\,dk_{(j+1)_{4}}...dp_{(j+1)_{4}}...\times \left\{ \gamma ^{\rho }\left[
\gamma ^{a}p_{(1)a}+..\right] ^{-1}..\left[ \gamma
^{b}p_{(I_{F})b}+..\right] ^{-1}\gamma _{\rho }\right.  \nonumber \\
&&\left[ k_{(1)}^{c}k_{(1)\,c}\right] ^{-1}..\left[
k_{(I_{B})}^{d}k_{(I_{B})\,d}\right] ^{-1}\times \left(
\prod_{l=1}^{3}N_{l}^{-1/4}|p_{(1)l}-p_{(2)l}+k_{(1)l}|^{-1}\right. 
\nonumber \\
&&\left. \left[ N_{l}^{-1/2}\delta \left( p_{(1)l}-p_{(2)l}+k_{(1)l}\right)
+i\,\mbox{sgn}\left( p_{(1)l}-p_{(2)l}+k_{(1)l}\right) \right] \right) 
\nonumber \\
&&+\mbox{similar terms}\,.  \label{eq:sixtwentyone}
\end{eqnarray}
The Dirac delta function $\delta \left( (\sum \left( E_{(l)}+\nu
_{(l)}\right) \right) $ indicates the conservation of total energy. The
number of integration variables in (\ref{eq:sixtwentyone}) is $%
3(I_{B}+I_{F})+(I_{B}+I_{F}-j+1)$. At high energy-momentum, the factor with
products of all $p^{-1},k^{-2},|p|^{-1}\mbox{sgn}(p)$ and $|k|^{-1}\,%
\mbox{sgn}(k)$ in the integrand, behaves asymptotically as the power $%
-I_{F}-2I_{B}-3j$. Therefore, counting the degree, $\kappa $, of
ultra-violet divergence with $p^{-1},k^{-2},|k|^{-1}\,\mbox{sgn}(k)$ and $%
|p|^{-1}\mbox{sgn}(p)$ factors alone, is 
\begin{equation}
\kappa :=3(I_{B}+I_{F})+(I_{B}+I_{F}-j+1)-I_{F}-2I_{B}-3j=1-E_{B}-(3/2)E_{F}.
\label{eq:sixtwentytwo}
\end{equation}
The degree, $\hat{\kappa}$ of divergence with product of all $|p|^{-1}\delta
(p)$ and $|k|^{-1}\delta (k)$ functions alone (after integrating out all $%
\delta $-functions) is 
\begin{equation}
\hat{\kappa}:=4(I_{B}+I_{F}-j+1)-I_{F}-2I_{B}-3j=4-3j-E_{B}-(3/2)E_{F}.
\label{eq:sixtwentythree}
\end{equation}
From (\ref{eq:sixtwentytwo}) and (\ref{eq:sixtwentythree}) we derive that 
\begin{equation}
\kappa -\widehat{\kappa }=3(j-1)\geq 0.  \label{eq:sixtwentyfour}
\end{equation}
There exist other degrees of divergences arising from the mixed products of $%
\mbox{sgn}(..)$ and $\delta (..)$. These degrees are all less or equal to $%
\kappa $. Therefore, the final criterion for ``convergence'' of the S-matrix
element $<f|\overline{S}_{j_{N}}|i>$ or $<f|\overline{S}_{j}|i>$ is that 
\begin{eqnarray}
\kappa &<&0,  \nonumber \\
(3/2)E_{F}+E_{B} &>&1.\   \label{eq:sixtwentyfive}
\end{eqnarray}
The above inequality proves that \emph{in case the number of external boson
lines is at least two, the S-matrix elements ``converge'' in every order of $%
j$}. However, in case the number of external bosonic lines is \emph{one},
the S-matrix element \emph{vanishes} by Furry's theorem. Therefore if there
is at least one bosonic external line, the S-matrix elements converge in
every order. Now, the number of incoming external fermionic lines always
equals to the number of outgoing fermionic lines. Therefore, we conclude
from the inequality (\ref{eq:sixtwentyfour}) that \emph{in case the number
of incoming external fermion lines is at least one, the S-matrix elements
``converge'' in every order $j$}. Thus, our final conclusion is that \emph{%
the S-matrix elements in the discrete phase space formulation converges in
every order provided there is at least one external line in the process}.

Let us consider some specific examples. In case of the self-energy of an
electron, as discussed in section-5, the relevant numbers are as follows: 
\begin{eqnarray}
E_{F} &=&2,\;\;\;\;E_{B}=0,\;\;\;\;\kappa =-2<0,  \nonumber \\
&&(3/2)E_{F}+E_{B}=3>1.  \label{eq:sixtwentysix}
\end{eqnarray}
Therefore, by the equations (\ref{eq:sixtwentyfour}), the S-matrix elements
representing the self-energy of the electron ``converges'' in every order.

As a second example, consider the self-energy of the photon. By the
treatment in section-5, we deal with the numbers: 
\begin{eqnarray}
E_{F} &=&0,\;\;\;\;E_{B}=2,\;\;\;\;\kappa =-1<0,  \nonumber \\
&&(3/2)E_{F}+E_{B}=2>1.  \label{eq:sixtwentyseven}
\end{eqnarray}
Again, the corresponding S-matrix elements for the self-energy of the photon
``converges'' in every order.

It can be shown that each of the so called \emph{primitive diagrams},
``converges'' in every order within the arena of the discrete phase space
formulation of the S-matrix.

\section{Acknowledgments}

\qquad I would like to express my gratitude to some people for their help in
different ways. I thank Prof. Y. Takahashi of Edmonton, Prof. A. Sen of
Allahbad and Prof. J. Gegenberg of Fredericton for informal discussions on
the quantum field theories. I would like to thank the great Prof. R. J.
Duffin of Pittsburgh for being my mentor in partial \emph{difference}
equations. I thank Prof. H. Srivastava of Victoria, Prof. G Malli and Dr. S.
Kloster of our university for their comments on the vertex function $\delta
^{\#}(p,q,k)$. I thank Dr. S. Das for improving my literary presentation. I
thank Mrs. J. Fabricus for typing papers I and II and Dr. A. DeBenedictis
for typing the paper-III.

\appendix

\section{Appendix: The distribution function $\delta _{3}^{\#}(\mathbf{p},%
\mathbf{q},\mathbf{k})$}

\qquad We define the distribution function $\delta^{\#}$ by the following
equation; 
\begin{eqnarray}
\delta^{\#}(p,q,k):=\sum_{n=0}^{\infty}\xi_{n}(p)\, \xi_{n}(q)\, \xi_{n}(k),
\nonumber \\
(p,q,k)\in \mathbf{R^3}.  \label{eq:a1}
\end{eqnarray}
We \emph{assume} that the right-hand side is at least summable in some
sense. The distribution function $\delta^{\#}$ so defined, is totally
symmetric with respect to the real variables $p,q,k$ and thus satisfies 
\begin{equation}
\delta^{\#}(p,q,k)=\delta^{\#}(q,p,k)=\delta^{\#}(p,k,q)= \delta^{\#}(k,q,p).
\label{eq:a2}
\end{equation}
It has also the following properties by the consequences (A.7) of paper-II: 
\begin{eqnarray}
\delta^{\#}(p,q,k)=\delta^{\#}(p,-q,-k)=
\delta^{\#}(-p,-q,k)=\delta^{\#}(-p,q,-k),  \label{eq:a3} \\
\overline{\delta^{\#}(-p,-q,-k)}=\delta^{\#}(p,q,-k)=\delta^{\#}(-p,q,k)=
\delta^{\#}(p,-q,k),  \label{eq:a4}
\end{eqnarray}

\qquad Now we shall state and prove in a simple manner the following theorem.

\textbf{Theorem A.1}: \emph{The distribution function} $\delta ^{\#}$ \emph{%
is distinct from the Dirac} $(2\pi )\delta $ \emph{-distribution}. \newline
\emph{Proof: Assuming} the interchangeability of the summation and
integration, and using the equations (II-A.I.1) and (II-A.I.19) we obtain 
\begin{eqnarray}
\int_{\mathbf{R}}\overline{\xi _{m}(k)}\delta ^{\#}(p,q,k)\,dk
&=&\sum_{n=0}^{\infty }\left[ \int_{\mathbf{R}}\xi _{m}(k)\xi
_{n}(k)\,dk\right] \xi _{n}(p)\xi _{n}(q)  \nonumber \\
&=&\xi _{m}(p)\xi _{m}(q).  \label{eq:a5}
\end{eqnarray}
(Here the subscript $m$ is not summed!) However, by the equation
(II-A.I.16), we have 
\begin{eqnarray}
\int_{\mathbf{R}}\overline{\xi _{m}(k)}\delta (p+q+k)dk &=&\overline{\xi
_{m}(-p-q)}=\xi _{m}(p+q)  \nonumber \\
&=&(\pi )^{4}2^{-m/2}\exp [(p-q)^{2}/2]  \nonumber \\
&&\times \sum_{j=0}^{m}\sqrt{\left( _{j}^{m}\right) }\xi _{m-j}(\sqrt{2}%
p)\xi _{j}(\sqrt{2}q).  \label{eq:a6}
\end{eqnarray}
Comparing the equations (\ref{eq:a5}) and (\ref{eq:a6}) we conclude that $%
\delta ^{\#}(p,q,k)\neq (2\pi )\delta (p+q+k)\Box $.

\qquad Now we shall introduce another sequence of functions 
\begin{eqnarray}
\zeta _{n}(k) &:&=\xi _{n}(k)\hbox{ for  }n\in \{0,1\},  \nonumber \\
\pi ^{-1/2}n^{-1/4}\cos (2k\sqrt{n})\hbox{  for }n &\in &\{2,4,6,...\}, 
\nonumber \\
i\pi ^{-1/2}n^{-1/4}\sin (2k\sqrt{n})\hbox{ for }n &\in &\{3,5,7,...\}.
\label{eq:a7}
\end{eqnarray}
If we introduce a new inner-product and norm for a non-separable space$%
^{8}:l_{2}^{\#}$ by 
\begin{eqnarray}
\langle \mathbf{\xi }|\mathbf{\eta }\rangle ^{\#} &:&=\lim_{N\rightarrow
\infty }\left[ (\pi /2\sqrt{N})\sum_{n=0}^{2N+1}\overline{\xi _{n}}\eta
_{n}\right] ,  \nonumber \\
||\mathbf{\xi }||^{\#} &:&=\sqrt{\langle \mathbf{\xi }|\mathbf{\xi }\rangle }%
,  \label{eq:a8}
\end{eqnarray}
then it can be shown by (\ref{eq:a7}) that 
\begin{equation}
||\mathbf{\xi (k)}-\mathbf{\zeta (k)}||\equiv 0.  \label{eq:a9}
\end{equation}
Therefore, it is reasonable to introduce an asymptotic approximation of $%
\delta ^{\#}(p,q,k)$ in (\ref{eq:a1}) by the expression 
\begin{eqnarray}
d^{\#}(p,q,k) &:&=\sum_{n=0}^{\infty }\zeta _{n}(p)\zeta _{n}(q)\zeta _{n}(k)
\nonumber \\
&\thickapprox &\pi ^{-3/4}e^{-(p^{2}+q^{2}+k^{2})/2}\left[
1-i\,2^{3/2}\,pqk\right]  \nonumber \\
&&+(1/8\pi ^{3/2})\sum_{n=2}^{\infty }n^{-3/4}\left[ e^{i2\sqrt{n}%
(p+q+k)}+e^{-i\,2\sqrt{n}(q+k-p)}\right.  \nonumber \\
&&+\left. e^{-i\,2\sqrt{n}(q-k+p)}+e^{i\,2\sqrt{n}(q-k-p)}\right] .
\label{eq:a10}
\end{eqnarray}
The convergence or summability of the above series is not obvious. However,
we shall discuss some related results. We can cite$^{9}$ the following
theorem. \newline

\textbf{Theorem A.2:} \emph{Let} $f$ \emph{and} $p$ \emph{be two real-valued
and continuous functions defined over} [a, $\infty $). \emph{Let} $%
\{b_{m}\}_{0}^{\infty }$ \emph{be a monotone increasing sequence such that} $%
b_{0}=a$ \emph{and} $\lim_{m\rightarrow \infty }b_{m}\rightarrow \infty $. 
\emph{Moreover, let} $f(x)>0$ \emph{and} $(-1)^{m}p(x)\geq 0$ \emph{for} $%
x\in [b_{m},b_{m+1}]\subset [a,\infty )$ \emph{Then, the improper (Riemann
or Lebesgue) integral} $\int_{a}^{\infty }f(x)p(x)\,dx$ \emph{converges to} $%
\alpha $ \emph{iff the alternating series} $\sum_{n=0}^{\infty
}c_{n}:=\sum_{n=0}^{\infty }\left[ \int_{b_{n}}^{b_{n+1}}f(x)p(x)\,dx\right] 
$ \emph{converges to} $\alpha $. (For proof see the reference 9.) Now we
shall prove the following corollary. \newline

\textbf{Corollary A.1:} \emph{There exists a sequence} $\lbrace
\Theta_{n}(k)\rbrace_{0}^{\infty}$ \emph{with the property} $\Theta_{n}(k)
\in (0, 1)$ \emph{for every} $n\in \mathbf{N}$ \emph{such that the series} $%
\sum_{n=0}^{\infty}\frac{\sin[(n+\Theta_{n}(k))\pi]}{[n+ \Theta_{n}(k)]^\beta%
}$ \emph{converges to} $\frac{\Gamma(1-\beta)\,\sin[(1-\beta)\pi/2]}{%
(k)^{1-\beta}}$ \emph{for all} $k>0$ \emph{and all} $\beta\in (0,1)$. 
\newline
\emph{Proof:} It is known$^{10}$ that for $k>0$ and $0<\beta<1$, the
improper integral 
\begin{eqnarray}
\int_{0^+}^{\infty}\frac{\sin(kx)}{x^\beta}\,dx = \frac{\Gamma(1-\beta) \sin[%
(1-\beta)\pi/2]}{(k)^{1-\beta}}  \nonumber \\
=\frac{\pi}{2(k)^{1-\beta}\cos[(1-\beta)\pi/2]\Gamma(\beta)} .
\label{eq:a11}
\end{eqnarray}
If we choose $a=0$, $f(x)=x^{-\beta}$, $p(x)=\sin(kx)\,\,\,(k>0)$, $%
b_{n}=(n\pi/k)$, and apply the preceding theorem-A.2, then we obtain that 
\begin{equation}
\frac{\Gamma(1-\beta)\,\sin[(1-\beta)\pi/2]}{(k)^{1-\beta}}=
\sum_{n=0}^{\infty} \left[\int_{n\pi/k}^{(n+1)\pi/k} \frac{\sin(kx)}{x^\beta}%
\,dx\right].  \nonumber
\end{equation}
Applying the mean-value theorem for an integral, we conclude that 
\begin{equation}
\int_{n\pi/k}^{(n+1)\pi/k} \frac{\sin(kx)}{x^\beta}= \frac{\sin[%
(n+\Theta_{n}(k))\pi]}{[n+\Theta_{n}(k)]^\beta} \;\;\;\hbox{ for
some  }\Theta_{n}(k)\in(0,1). \Box  \nonumber
\end{equation}

\qquad We can generalize the preceding corollary for any $k\neq 0$ and $%
\beta \in (0,1)$ to the following equations: 
\begin{eqnarray}
\sum_{n=0}^{\infty }\frac{\sin [(n+\Theta _{n}(k))\pi ]}{[n+\Theta
_{n}(k)]^{\beta }} &=&\int_{0^{+}}^{\infty }\frac{\sin (kx)}{x^{\beta }}\,dx
\nonumber \\
&=&\frac{\hbox{sgn}(k)\,\,\Gamma (1-\beta )\,\,\sin [(1-\beta )\pi /2]}{%
|k|^{1-\beta }},  \label{eq:a12}
\end{eqnarray}
\begin{eqnarray}
\sum_{n=0}^{\infty }\frac{\cos [(n+\hat{\Theta}_{n}(k))\pi ]}{[n+\hat{\Theta}%
_{n}(k)\pi ]^{\beta }} &=&\int_{0^{+}}^{\infty }\frac{\cos (kx)}{x^{\beta }}%
\,dx  \nonumber \\
&=&\cos [(1-\beta )\pi /2]\{\frac{\Gamma (1-\beta )}{|k|^{1-\beta }}-2\Gamma
(-\beta )|k|^{\beta }\delta (k)\},  \nonumber
\end{eqnarray}
\begin{eqnarray}
\sum_{n=0}^{\infty }\frac{\exp \{i[(n+\Theta _{n}^{\#}(k))\pi ]\}}{[n+\Theta
_{n}^{\#}(k)\pi ]^{\beta }} &=&\int_{0^{+}}^{\infty }\frac{e^{ikx}}{x^{\beta
}}\,dx  \nonumber \\
&=&\frac{\Gamma (1-\beta )}{|k|^{1-\beta }}\exp \{i\hbox{ sgn}(k)(1-\beta
)\pi /2\} \\
&&-2\cos [(1-\beta )\pi /2]\Gamma (-\beta )|k|^{\beta }\delta (k).  \nonumber
\end{eqnarray}
Here, $\delta (k)$ is the Dirac delta distribution.

\qquad It is very plausible from (\ref{eq:a12}iii) that the series $%
d^{\#}(p,q,k)$ in (\ref{eq:a10}) converge. In fact, from (\ref{eq:a12}i, ii,
iii) we shall choose an approximation for $d^{\#}(p,q,k)$ by the following
relation: 
\begin{eqnarray}
&&8d^{\#}(p,q,k)\approx \pi ^{-3/2}\int_{0^{+}}^{\infty }y^{-3/4}\left( \exp
\{i[2(p+q+k)\sqrt{y}]\}\right.  \nonumber \\
&&+\left. \exp \{-i[2(q+k-p)\sqrt{y}]\}+\exp \{-i[2(q-k+p)\sqrt{y}]\}+\exp
\{i[2(q-k-p)\sqrt{y}]\}\right) \,dy  \nonumber \\
&=&\sqrt{2}(\pi )^{-3/2}\int_{0^{+}}^{\infty }x^{-1/2}\{\exp [i(p+q+k)x] 
\nonumber \\
&&+\exp [-i(q+k-p)x]+\exp [-i(q-k+p)x]+\exp [i(q-k-p)x]\}dx.  \label{eq:a13}
\end{eqnarray}

\qquad Now, choosing the special case $\beta =1/2$ and recalling $\Gamma
(1/2)=\sqrt{\pi },\,\,\Gamma (-1/2)=-2\sqrt{\pi }$, we obtain from (A.12iii) 
\begin{equation}
\int_{0^{+}}^{\infty }x^{-1/2}e^{ikx}\,dx=\sqrt{\pi }\{2\sqrt{2|k|}\delta
(k)+\frac{[1+i\hbox{ sgn}(k)]}{\sqrt{2|k|}}\}.  \label{eq:a14}
\end{equation}
Substituting (\ref{eq:a14}) into (\ref{eq:a13}), we finally obtain an
asymptotically ``closed form'': 
\begin{eqnarray}
2\pi \,d^{\#}(p,q,k) &=&\{\sqrt{|p+q+k|}\delta (p+q+k)+\sqrt{|q+k-p|}\delta
(q+k-p)  \nonumber \\
&+&\sqrt{|q-k+p|}\delta (q-k+p)+\sqrt{|q-k-p|}\delta (q-k-p)\}.  \nonumber \\
&+&(1/4)\{\frac{[1+i\hbox{ sgn}(p+q+k)]}{\sqrt{|p+q+k|}}+\frac{[1-i%
\hbox{
sgn}(q+k-p)]}{\sqrt{|q+k-p|}}  \nonumber \\
&+&\frac{[1-i\hbox{ sgn}(q-k+p)]}{\sqrt{|q-k+p|}}+\frac{[1+i\hbox{ sgn}%
(q-k-p)]}{\sqrt{|q-k-p|}}\}.  \label{eq:a15}
\end{eqnarray}

\qquad We shall now state and sketch briefly the proof of the \emph{%
remarkable properties} of the distribution function $d^{\#}(p,q,k)$ in (\ref
{eq:a15}). \newline

\textbf{Theorem A.3:} \emph{The distribution function} $d^{\#}(p,q,k)$ \emph{%
satisfies exactly} \emph{the symmetry properties} (A.\ref{eq:a2}, \ref{eq:a3}%
, \ref{eq:a4}) \emph{of} $\delta ^{\#}(p,q,k)$.\smallskip

\noindent \emph{Proof:} The proof is straight forward (using \ref{eq:a15})
and is skipped. \newline

\qquad We define the three dimensional distribution function 
\begin{equation}
d_{3}^{\#}(\mathbf{p,q,k}):=\prod_{j=1}^{3}d^{\#}(p_{j},q_{j},k_{j}).
\label{eq:a17}
\end{equation}

\qquad In section-V, we shall use the \emph{asymptotic approximation} of the 
$\delta_{3}^{\#}$-distribution function by putting 
\begin{equation}
\delta_{3}^{\#}(\mathbf{p, q, k})\approx d_{3}^{\#}(\mathbf{p, q, k}) .
\label{eq:a18}
\end{equation}

\newpage

\section{Bibliography}

$^1$ L. Gross, \emph{Comm. Pure and Appl. Math}. \textbf{19}, 1 (1966); 
\newline
J. Chadam, \emph{J. Math. Phys}. \textbf{13}, 597 (1972); \newline
M. Flato, J. C. H. Simon, and E. Taflin, \emph{Comm. Math. Phys}. \textbf{112%
}, 1 (1987); \newline
A. Das and D. Kay, \emph{J. Math. Phys}. \textbf{30}, 2280 (1989); \newline
A. Das, \emph{J. Math. Phys}. \textbf{34}, 3986 (1993); \newline
H. S. Booth and C. J. Radford, \emph{J. Math. Phys}. \textbf{38}, 1257
(1997). \newline
\newline
$^2$ A. Das and P. Smoczynski, \emph{Found. Phys. Lett.} \textbf{7}, 127
(1994). \newline
\newline
$^3$ N. N. Bogolubov and D. V. Shirkov, \emph{Introduction to the theory of
Quantized Fields} (Translated by G. M. Volkoff) (Interscience Publishers
Inc., New York, 1959), p.648. \newline
\newline
$^4$ A. Das, \emph{Nuovo Cimento}, \textbf{18}, 482 (1960). \newline
\newline
$^5$ H. Muirhead, \emph{The Physics of Elementary Particles} (Pergamon
Press, Oxford, 1965), pp.290, 313. \newline
\newline
$^6$ S. Weinberg, \emph{The Quantum Theory of Fields Vol. I} (Cambridge
University Press, Cambridge, 1995), pp.114, 175. \newline
\newline
$^7$ M. E. Peskin and D. V. Schroeder, \emph{An Introduction to Quantum
Field Theory} (Addison - Wesley Publishing, Reading Massachusetts, 1995),
p.89. \newline
\newline
$^8$ A. Das and P. Smoczynski, \emph{Found. Phys. Lett.}, \textbf{7}, 21
(1994). \newline
\newline
$^9$ W. Kaplan, \emph{Advanced Calculus} (Addison - Weseley Publishing Co.
Inc., Reading, Massachusetts, 1959), p.374. \newline
\newline
$^{10}$ I. S. Gradshteyn and I. M. Ryzhik, \emph{Tables of Integrals,
Series, and Products} (Academic Press Inc., San Diego, 1980), p.420.

\end{document}